\documentclass[twocolumn]{aastex701}
\usepackage{amsmath, amssymb}
\usepackage{CJK}

\defcitealias{Davis2025}{D25}
\begin{document}
\begin{CJK*}{UTF8}{gbsn}
\title{Summary of the First Year of the Space Weather Around Young Suns Program: 900 Hours of Low-frequency Radio and Optical Data Dedicated to Young, Solar-type Stars}

\author[0000-0001-5397-5969]{Ivey Davis}
\affiliation{Cahill Center for Astronomy and Astrophysics, California Institute of Technology, Pasadena, CA 91125, USA}
\affiliation{Owens Valley Radio Observatory, California Institute of Technology, Big Pine, CA 93513, USA}
\affiliation{ASTRON, Netherlands Institute for Radio Astronomy, Oude Hoogeveensedijk 4, Dwingeloo, 7991 PD, The Netherlands}
\email[show]{idavis@caltech.edu}  
\author[0000-0002-7083-4049]{Gregg Hallinan}
\affiliation{Cahill Center for Astronomy and Astrophysics, California Institute of Technology, Pasadena, CA 91125, USA}
\affiliation{Owens Valley Radio Observatory, California Institute of Technology, Big Pine, CA 93513, USA}
\email{}
\author[0000-0003-1226-118X]{Nikita Kosogorov}
\affiliation{Cahill Center for Astronomy and Astrophysics, California Institute of Technology, Pasadena, CA 91125, USA}
\affiliation{Owens Valley Radio Observatory, California Institute of Technology, Big Pine, CA 93513, USA}
\email{}
\author{Marin M. Anderson}
\affiliation{Owens Valley Radio Observatory, California Institute of Technology, Big Pine, CA 93513, USA}
\affiliation{Jet Propulsion Laboratory, California Institute of Technology, Pasadena, CA 91011, USA}
\email{}
\author{John Baker}
\affiliation{Caltech Optical Observatories, California Institute of Technology, Pasadena, CA 91125, USA}
\email{}
\author{Judd D. Bowman}
\affiliation{School of Earth and Space Exploration, Arizona State University, Tempe, AZ 85287, USA}
\email{}
\author{Rick Burruss}
\affiliation{Caltech Optical Observatories, California Institute of Technology, Pasadena, CA 91125, USA}
\email{}
\author{Ruby Byrne}
\affiliation{Cahill Center for Astronomy and Astrophysics, California Institute of Technology, Pasadena, CA 91125, USA}
\affiliation{Owens Valley Radio Observatory, California Institute of Technology, Big Pine, CA 93513, USA}
\email{}
\author{Morgan Catha}
\affiliation{Owens Valley Radio Observatory, California Institute of Technology, Big Pine, CA 93513, USA}
\email{}
\author{Bin Chen}
\affiliation{Center for Solar-Terrestrial Research, New Jersey Institute of Technology, Newark, NJ 07102, USA}
\email{}
\author{Xingyao Chen}
\affiliation{Center for Solar-Terrestrial Research, New Jersey Institute of Technology, Newark, NJ 07102, USA}
\email{}
\author[0000-0001-7754-0804]{Sherry Chhabra}
\affiliation{George Mason University, Fairfax, VA 22030, USA}
\email{}
\author{Curt Corcoran}
\affiliation{Caltech Optical Observatories, California Institute of Technology, Pasadena, CA 91125, USA}
\email{}
\author{Larry D'Addario}
\affiliation{Cahill Center for Astronomy and Astrophysics, California Institute of Technology, Pasadena, CA 91125, USA}
\affiliation{Owens Valley Radio Observatory, California Institute of Technology, Big Pine, CA 93513, USA}
\email{}
\author{Jayce Dowell}
\affiliation{University of New Mexico, Albuquerque, NM 87131, USA}
\email{}
\author{Katherine Elder}
\affiliation{School of Earth and Space Exploration, Arizona State University, Tempe, AZ 85287, USA}
\email{}
\author{Dale Gary}
\affiliation{Center for Solar-Terrestrial Research, New Jersey Institute of Technology, Newark, NJ 07102, USA}
\email{}

\author{Charlie Harnach}
\affiliation{Owens Valley Radio Observatory, California Institute of Technology, Big Pine, CA 93513, USA}
\email{}
\author{Carolyn Heffner}
\affiliation{Caltech Optical Observatories, California Institute of Technology, Pasadena, CA 91125, USA}
\email{}
\author{Greg Hellbourg}
\affiliation{Cahill Center for Astronomy and Astrophysics, California Institute of Technology, Pasadena, CA 91125, USA}
\affiliation{Owens Valley Radio Observatory, California Institute of Technology, Big Pine, CA 93513, USA}
\email{}
\author{Jack Hickish}
\affiliation{Real-Time Radio Systems Ltd, Bournemouth, Dorset BH6 3LU, UK}
\email{}
\author{Rick Hobbs}
\affiliation{Owens Valley Radio Observatory, California Institute of Technology, Big Pine, CA 93513, USA}
\email{}
\author{David Hodge}
\affiliation{Cahill Center for Astronomy and Astrophysics, California Institute of Technology, Pasadena, CA 91125, USA}
\email{}
\author{Mark Hodges}
\affiliation{Owens Valley Radio Observatory, California Institute of Technology, Big Pine, CA 93513, USA}
\email{}
\author{Yuping Huang}
\affiliation{Cahill Center for Astronomy and Astrophysics, California Institute of Technology, Pasadena, CA 91125, USA}
\affiliation{Owens Valley Radio Observatory, California Institute of Technology, Big Pine, CA 93513, USA}
\email{}
\author{Andrea Isella}
\affiliation{Department of Physics and Astronomy, Rice University, Houston, TX 77005, USA}
\email{}
\author{Daniel C. Jacobs}
\affiliation{School of Earth and Space Exploration, Arizona State University, Tempe, AZ 85287, USA}
\email{}
\author{Ghislain Kemby}
\affiliation{Owens Valley Radio Observatory, California Institute of Technology, Big Pine, CA 93513, USA}
\email{}
\author{John T. Klinefelter}
\affiliation{Owens Valley Radio Observatory, California Institute of Technology, Big Pine, CA 93513, USA}
\email{}
\author{Matthew Kolopanis}
\affiliation{School of Earth and Space Exploration, Arizona State University, Tempe, AZ 85287, USA}
\email{}
\author{James Lamb}
\affiliation{Owens Valley Radio Observatory, California Institute of Technology, Big Pine, CA 93513, USA}
\email{}
\author{Casey Law}
\affiliation{Cahill Center for Astronomy and Astrophysics, California Institute of Technology, Pasadena, CA 91125, USA}
\affiliation{Owens Valley Radio Observatory, California Institute of Technology, Big Pine, CA 93513, USA}
\email{}
\author{Nivedita Mahesh}
\affiliation{Cahill Center for Astronomy and Astrophysics, California Institute of Technology, Pasadena, CA 91125, USA}
\affiliation{Owens Valley Radio Observatory, California Institute of Technology, Big Pine, CA 93513, USA}
\email{}
\author[0000-0002-2325-5298]{Surajit Mondal}
\affiliation{Center for Solar-Terrestrial Research, New Jersey Institute of Technology, Newark, NJ 07102, USA}
\email{}
\author{Navtej Saini}
\affiliation{Jet Propulsion Laboratory, California Institute of Technology, Pasadena, CA 91011, USA}
\email{}
\author{Brian O'Donnell}
\affiliation{Center for Solar-Terrestrial Research, New Jersey Institute of Technology, Newark, NJ 07102, USA}
\email{}
\author{Kathryn Plant}
\affiliation{Owens Valley Radio Observatory, California Institute of Technology, Big Pine, CA 93513, USA}
\affiliation{Jet Propulsion Laboratory, California Institute of Technology, Pasadena, CA 91011, USA}
\email{}
\author{Corey Posner}
\affiliation{Owens Valley Radio Observatory, California Institute of Technology, Big Pine, CA 93513, USA}
\email{}
\author{Travis Powell}
\affiliation{Owens Valley Radio Observatory, California Institute of Technology, Big Pine, CA 93513, USA}
\email{}
\author{Vinand Prayag}
\affiliation{Owens Valley Radio Observatory, California Institute of Technology, Big Pine, CA 93513, USA}
\email{}
\author{Andres Rizo}
\affiliation{Owens Valley Radio Observatory, California Institute of Technology, Big Pine, CA 93513, USA}
\email{}
\author{Andrew Romero-Wolf}
\affiliation{Jet Propulsion Laboratory, California Institute of Technology, Pasadena, CA 91011, USA}
\email{}
\author{Jun Shi}
\affiliation{Cahill Center for Astronomy and Astrophysics, California Institute of Technology, Pasadena, CA 91125, USA}
\email{}
\author{Greg Taylor}
\affiliation{University of New Mexico, Albuquerque, NM 87131, USA}
\email{}
\author{Jordan Trim}
\affiliation{Owens Valley Radio Observatory, California Institute of Technology, Big Pine, CA 93513, USA}
\email{}
\author{Mike Virgin}
\affiliation{Owens Valley Radio Observatory, California Institute of Technology, Big Pine, CA 93513, USA}
\email{}
\author[0000-0002-6611-2668]{Akshatha K. Vydula}
\affiliation{School of Earth and Space Exploration, Arizona State University, Tempe, AZ 85287, USA}
\email{}
\author{Sandy Weinreb}
\affiliation{Cahill Center for Astronomy and Astrophysics, California Institute of Technology, Pasadena, CA 91125, USA}
\email{}
\author{Scott White}
\affiliation{Owens Valley Radio Observatory, California Institute of Technology, Big Pine, CA 93513, USA}
\email{}
\author{David Woody}
\affiliation{Owens Valley Radio Observatory, California Institute of Technology, Big Pine, CA 93513, USA}
\email{}
\author[0000-0003-2872-2614]{Sijie Yu}
\affiliation{Center for Solar-Terrestrial Research, New Jersey Institute of Technology, Newark, NJ 07102, USA}
\email{}
\author{Thomas Zentmeyer}
\affiliation{Owens Valley Radio Observatory, California Institute of Technology, Big Pine, CA 93513, USA}
\email{}
\author[0000-0001-6855-5799]{Peijin Zhang (张沛锦)}
\affiliation{Center for Solar-Terrestrial Research, New Jersey Institute of Technology, Newark, NJ 07102, USA}
\email{}
\author{Jeffry Zolkower}
\affiliation{Caltech Optical Observatories, California Institute of Technology, Pasadena, CA 91125, USA}
\email{}


\begin{abstract}
    The Space Weather Around Young Suns (SWAYS) program was introduced in \citet{Davis2025} as a multi-wavelength monitoring program for studying the activity and particle environments of nearby, young, solar-type stars. The SWAYS program currently includes the Owens Valley Radio Observatory Long Wavelength Array (OVRO-LWA) operating between 13--87\,MHz to search for stellar equivalents of solar type~II and III bursts, which are associated with bulk plasma motion in the corona and interplanetary medium. These observations are accompanied by simultaneous photometric data from the high-precision, optical instrument Flarescope to identify associated flare events.
    These two instruments have collectively acquired nearly 900\,hr of data with $\approx70\%$ overlap between November 2023--June 2024, dedicated to six stars.
    Here, we present the results of this first season of the SWAYS observing campaign, which include a superflare from the star EK~Draconis with no accompanying low-frequency particle-flux signal. The novelty of the coordination at these specific parts of the spectrum allow us to uniquely evaluate the conditions that may have inhibited a radio detection. We find that the exceptionally hot, dense coronae of incredibly active stars may not be conducive to the development of the instabilities required for type~II and III bursts, or else inspire new expectations for when we should expect to observe a signal relative to the time of the flare. This may represent the plasma-density complement to the magnetospheric limitations to observing space-weather signatures at low frequencies.
\end{abstract}

\section{Introduction}
We have inferred from studying young, solar-type stars that the Sun was likely much more magnetically active in its youth \citep{Wilson1963} and that this activity may have facilitated stronger winds \citep{Wood2005}.
The presence of high activity levels and winds has important consequences for the overall angular-momentum evolution of the Sun as well as the environment that the young Earth may have experienced.
In addition to the increased mass loss and, consequently, denser quasi-steady wind that the young Sun may have been producing, the higher levels of activity may have also facilitated an higher number of larger and faster coronal mass ejections (CMEs).
This expectation is inspired by scaling laws investigated by, e.g., \citet{Yashiro2006} and \citet{Yashiro2009}, which show that a CME from the Sun is essentially guaranteed for flare energies $>10^{30}\,$erg and that the energy, mass, and velocity of such solar CMEs are correlated with X-ray flare properties like the peak luminosity.

Some investigations have suggested that CMEs may have contributed even more to the overall mass loss than the ambient wind in the early solar history \citep{Cranmer2017CME}.
This would have incredible implications both for our distinction between a quasi-steady wind and individual events as well as for the extreme pressure and ion content contributed to the Earth and the chemistry that would follow.
Alternatively, the strong magnetic fields that would be required to drive such energetic events may also inhibit bulk mass escape \citep{AlvaradoGomez2018}, or at least inhibit signatures of such events \citep{Villadsen2019}.

Searching for signatures of stellar CMEs to determine their environmental impacts is difficult, and corroborating such signals as genuine indicators of mass \textit{escape} rather than simply mass \textit{motion} is even more so.
This is because our most successful methods for identifying mass motion from main sequence stars are Doppler spectroscopy \citep{Namekata2022Nature} and X-ray or extreme ultra-violet (EUV) dimming \citep{Veronig2021, Loyd2022}, which are limited to material relatively low in the stellar corona.
Some have explored the possibility of determining if a CME occurred based on its impact on exoplanetary signatures \citep{Lazio2004_BodesLaw, AlvaradoGomez2022}, which would imply mass escape far into the interplanetary medium.
An alternative solution is looking for plasma emission at low frequencies ($<1$\,GHz).
Such emission is responsible for the type~II and III bursts we observe from the Sun; the former are associated with shocks driven by prominences and CMEs and the latter are associated with other solar energetic particles (SEPs), including those accelerated along opened field lines. 
The plasma frequency, $\nu_{\text{p}}$, that the emission is observed at scales with the \textit{ambient} plasma density \added{like $ \nu_{\text{p}}\approx9\sqrt{n_{\text{e}^-}\text{cm}^{-3}}$\,kHz}---emission occurring at sufficiently low frequencies may be attributed to low-density material far from the stellar surface.
The dependence on the ambient density also leads to distinctive sweeping structures in dynamic spectra: the slow ($\lesssim1000\,$km\,s$^{-1}$), broad shock fronts of CMEs produce the broad-band, slowly-drifting type~II signatures in dynamic spectra \citep{NelsonMelrose1985} while the fast ($\gtrsim0.3\,c$, where $c$ is the speed of light) electron beams responsible for type~III bursts produce relatively narrow, fast-drifting emissions \citep{Dulk1985}. 

The Space Weather Around Young Suns (SWAYS) program was designed to identify plasma emission from nearby solar-type stars as well as the reconnection event responsible for accelerating the material \citep[][hereafter \citetalias{Davis2025}]{Davis2025}.
The program currently consists of the Owens Valley Radio Observatory Long Wavelength Array (OVRO-LWA), operating between 13--87\,MHz, which should be sensitive to extreme type~III bursts occurring within $\approx15\,$pc.
If solar scaling laws between X-ray flare luminosity and type~II burst luminosity hold for other stars, then this detection horizon may be much further \citep{Mohan2024_scaling_relation}.
The program also includes Flarescope, a high-precision, optical-photometry instrument located at Palomar mountain and which is sensitive to flares with bolometric energies $\gtrsim10^{33}\,$erg assuming a flare temperature $T_{\text{fl}}=10,000\,$K.
Here, we present the results of the first season of SWAYS operations, which  began in November of 2023 and paused in July 2024. 
In that time, Flarescope completed 500 hours of observing across six young, solar-type stars with 388 hours of complementary beamformed data from the OVRO-LWA, summarized in Table \ref{tab:stellar_info}.
This constitutes the longest dedicated monitoring program for solar-type stars $<100\,$MHz and the longest \added{targeted }radio monitoring program with complementary optical data.

In Section~\ref{sec:data}, we provide an overview of the SWAYS observing program, including brief summaries of the stellar sample and the data products from Flarescope and the OVRO-LWA.
In Section~\ref{sec:flare_search}, we review the details and results of the flare search in Flarescope data, which includes a detection of a $\sim10^{34}\,$erg superflare from \added{the highly active G-dwarf }EK~Draconis (EK~Dra). 
\added{This star has been studied robustly as a proxy for the young Sun \citep[e.g.,][]{Gudel1995, Telleschi2005, Gudel2007, Ayres2015, Fichtinger2017} and has had a substantial number of candidate CME detections \citep{Namekata2025V889Her}.}
We describe the reduction of the cross-correlated data from the OVRO-LWA for the night of this flare in Section~\ref{sec:ovro_lwa_data_reduction}. 
In Section \ref{sec:ovro_lwa_results} we describe the search for type~II and III burst candidates in this data; although the energy of the flare we observed has been associated with chromospheric signals of eruptive prominences on EK~Dra \citep{Namekata2022, Namekata2024}, here we find no escape signatures originating from high in EK~Dra's corona.
We discuss the significance of the non-detection of low-frequency bursts in Section~\ref{sec:discussion}, where we investigate the possibility of the high temperature and density of EK~Dra's corona inhibiting the development of bursts.

\begin{deluxetable*}{l c c c c c c c c c c c}
\tablecaption{Properties and completed observing time for the six main stars apart of the SWAYS sample. }\label{tab:stellar_info}
\tablehead{
\colhead{Star}  & \colhead{RA} & \colhead{Dec} & \colhead{$d$} & \colhead{$T_q$} & \colhead{$R_\star$}&\colhead{Flare Detection Threshold} & \colhead{Burst Detection Threshold} & \colhead{Flarescope} & \colhead{OVRO-LWA} & \colhead{Overlap} \\
          \colhead{} & \colhead{[$^\circ$]} & \colhead{[$^\circ$]} & \colhead{[pc]} & \colhead{[K]} & \colhead{[$R_\odot$]}&\colhead{[erg]} & \colhead{[SFU]} & \colhead{hours} & \colhead{hours} & \colhead{}
           }
        \startdata
         $\epsilon$ Eridani & 53.23&-9.46& 3.21 & 5002\textsuperscript{G22}& 0.83& $2.3 \times10^{32}$& $3.3\times10^7$& 153 & 97&51\%\\
         $\chi^1$\,Orionis & 88.59&+20.28& 8.66 & 5808\textsuperscript{G22}& 1.5& $1.78 \times10^{33}$& $2.4\times10^8$& 19& 20&70\%\\
         $\kappa^1$ Ceti & 49.84&+3.37& 9.14 & 5694\textsuperscript{P}& 0.96& $6.6 \times10^{32}$  &  $2.7\times10^8$&  2& 4&44\%\\
         $\pi^1$\,Ursa Majoris  & 129.80&+65.02& 14.4 & 5711\textsuperscript{G22}& 0.98& $7.0 \times10^{32}$ & $6.6\times10^8$& 95& 69&71\%\\
         EK Draconis  & 219.75&+64.29& 34.4 &  5423\textsuperscript{G22}& 1.01& $5.4 \times10^{32}$ &$3.8\times10^9$ & 189& 168&76\%\\
        V889 Herculis & 278.58& +18.69& 35.6&  5516\textsuperscript{G22}& 1.06& $6.7 \times10^{32}$&$4.0\times10^9$ & 42& 30&71\%\\
        \enddata
\tablecomments{Radii $R_\star$ come from \citet{GaiaCollaboration2018} and distances $d$ come from \citet{Gaia2022}. The flare detection threshold is determined assuming a $5\sigma=2.5\,$mmag measurement in a single, $90\,$s integration and a flare temperature of 10,000\,K. The burst detection threshold is calculated assuming a $5\sigma_{\text{th}}=750\,$mJy, where $\sigma_{\text{th}}$ is the theoretical, full-band cross-correlated sensitivity; it is reported in solar flux units (SFU). All observations were conducted in the Sloan $g'$ band with either a $1^\circ$ or $0.25^\circ$ diffuser on Flarescope and with a 256\,ms integration time for the OVRO-LWA beamformer. The OVRO-LWA time is calculated based on the number of hours spent on each star with the beamformer; there is some additional overlap to be accounted for from the cross-correlated data.}
\tablerefs{(G22) \citet{Gaia2022}; (P) \citet{Pal2023}}
\end{deluxetable*}

\section{Data}\label{sec:data}
The coordination of the OVRO-LWA and Flarescope in the SWAYS program is described in detail in \citetalias{Davis2025}.
Here, we briefly review the stellar sample, relevant data products, and general observing details for the program.

\subsection{Stellar Sample}
The original SWAYS program contained five stars---$\epsilon$~Eridani (eps Eri), $\chi^1$~Orionis (chi Ori), $\kappa^1$~Ceti (kap Cet), $\pi^1$~Ursa Majoris (pi UMa), and EK Dra.
These are all active, solar-type stars (effective photospheric temperatures $T_{\text{eff}}\approx5000$--$6000\,$K and surface gravity $\log g \gtrsim4$) with ages spanning $\sim0.1$--$1\,$Gyr.
The latter four stars have been popular objects for investigating solar history and evolution due to their youth and the similarity of their spectral properties to the Sun's, with each star being used to represent a snapshot of the Sun during its first $\approx1\,$Gyr on the main sequence \citep{Dorren_Guinan_1994, Gudel1997, Telleschi2005, Gudel2007, Fichtinger2017, Leitzinger2020}.
eps~Eri also has been popular to study due to its youth and close proximity.
As the nearest solar-type star to host an exoplanet \citep{Mawet2019}, the possibility of CMEs occurring on eps~Eri make it a promising testbed for studying exo-solar space weather \citep{Loyd2022}. 

The star V889~Herculis (V889~Her) was added to the sample near the end of the observing season.
This star is estimated to be less than $50\,$Myr and thus has not quite yet entered the main sequence.
There is a significant amount of research dedicated to V889~Her's magnetic field in the context of its long-lived starspots \citep{Strassmeier2003, Marsden2006, Jarvinen2008, Willamo2022field_rotation}. 
However, it has only recently been investigated rigorously in the context of flare activity \citep{Namekata2025V889Her}.
Due to V889~Her's high activity levels and the extensive research that has been dedicated to understanding its field, we added it to the SWAYS sample as a star to prospectively represent the Sun just before it turned onto the main sequence.
Relevant properties for these stars, including event detection thresholds for Flarescope and the OVRO-LWA, and the number of hours spent on each star by each instrument are summarized in Table \ref{tab:stellar_info}.

\subsection{Flarescope}\label{sec:data.flarescope}
Flarescope is a 0.5\,m-aperture, optical photometry instrument on Palomar Mountain. 
It was designed explicitly to detect flare events on nearby, solar-type stars during coordinated observations with the OVRO-LWA.
The high optical fluxes from these stars, coupled with the low contrast of flare events against the quiescent solar-type photospheres, required that Flarescope be \added{capable of} sub-milli-mag (sub-mmag) precision without saturating its detector.
To meet these requirements, Flarescope uses a Sloan $g'$ filter (400--550\,nm), engineered diffusers that provide a top-hat point-spread function (PSF), and a frame-transfer imaging mode \added{that facilitates both} low read noise \added{and a high} imaging cadence. 

When operating with the Sloan $g'$ filter and a $1^\circ$ diffuser (diffusing over $\approx1700\,$pix on the detector), Flarescope is capable of reaching sub-mmag precision at $1\,$min integration time for pi~UMa (Sloan $g'$ magnitude $=5.5$).
Because scintillation is the dominant noise factor for the SWAYS targets, Flarescope should be capable of reaching similar precision on the same timescale for all SWAYS targets, assuming the observing conditions are the same quality. 
All SWAYS observations were conducted in the Sloan $g'$ band with either the $1^\circ$ or $0.25^\circ$ diffuser.
Exposure time during the observing campaign varied by source between 1.1--5\,s depending on the flux received per pixel and \added{the }diffuser setting, with 1.1\,s being the shortest exposure time possible when using the frame-transfer mode of the detector. 
We follow the data-reduction and light-curve extraction processes outlined in \citetalias{Davis2025} and perform our flare search on light curves at 5\,s and 90\,s integration times. 
The long integration time is informed by the timescale needed to reach sub-mmag precision on pi~UMa and the short integration time is to accommodate short, low-amplitude flares that may be diluted at longer integration times.

\subsection{OVRO-LWA}\label{sec:data.lwa}
The OVRO-LWA is a 352-element, cross-dipole interferometer operating between 13--87\,MHz with 24\,kHz spectral resolution. 
241 of these elements are contained within 200\,m to provide dense \textit{uv}-plane coverage; these elements constitute the core. 
An additional 111 antennas act as outriggers, providing a maximum baseline of 2.3\,km.
The recently-upgraded backend supports multiple simultaneous observing modes. 
The two modes that are relevant for this work are the beamformed data produced from the incoherent sum of antenna powers and the cross-correlated data.

The cross-correlated data provide all-sky snapshot images integrated every 10\,s and are collected continuously all day, every day. 
The spatial information, capacity for direction-dependent calibration, and means for sky subtraction make this mode ideal for searching for burst candidates. 
However, the coarse time resolution may not be suitable for resolving type~III burst structures, which may drift through the band in a single integration.
Alternatively, beamformed data suffers from integrating a substantial amount of the sky and being limited in doing nuanced radio-frequency interference (RFI) flagging. 
That said, it has high time resolution (up to one millisecond) that may be essential for characterizing bursts. 
Employing a combination of background-subtraction techniques along with the incoherent de-dispersion method described in \citetalias{Davis2025} may be sufficient to reveal potential signals.
 For the SWAYS observing program, a power beam tracked the same stars which Flarescope observed on a given night and collected data at 256\,ms time resolution. 
Cross-correlated data were collected simultaneously with these observations for many of the nights.

Because radio data---especially cross-correlated data---are computationally expensive to process and analyze, the OVRO-LWA datasets that we search for bursts from are prioritized based on the presence of flares in the Flarescope data.
In the case that a flare is observed by Flarescope, we investigate the images produced from cross-correlated data for evidence of a burst at the location of the star. 
If there is a signal in the cross-correlated images, we would then follow up the search in the high-time-resolution beamformed data to characterize the burst structure. 

\section{Flare search}\label{sec:flare_search}
\begin{figure*}
    \centering
    \includegraphics[width=\textwidth]{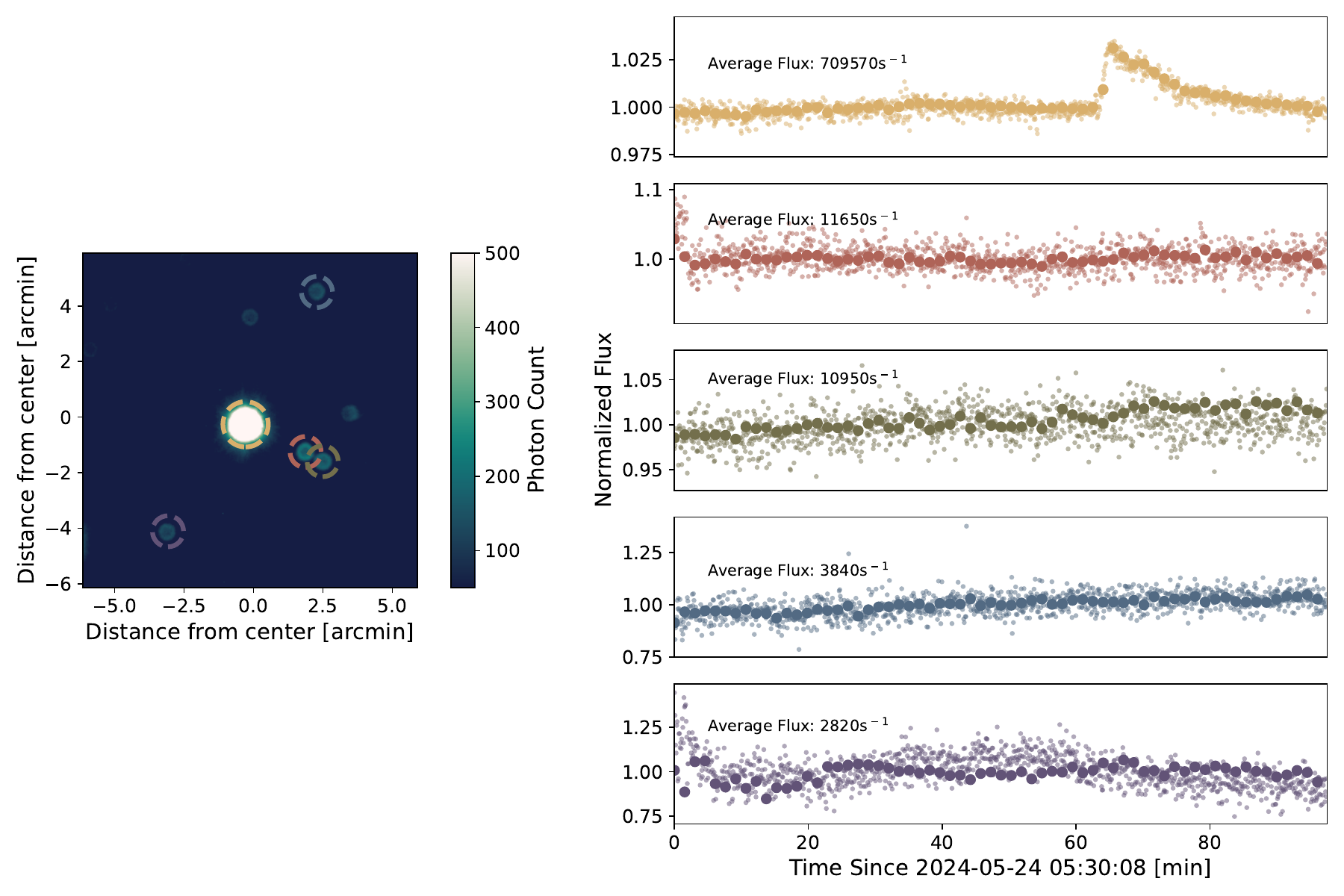}
    \caption{The FOV for EK Dra (left) and the flux-normalized light curves for EK Dra and the four other brightest stars in the FOV (right panels). The light curves' colors correspond to the color of the dashed-line aperture in the field, small scatter points are the 5\,s light curves, and large scatter points are the 90\,s light curves. The distinctive steep rise and exponential decay of a flare is apparent in EK Dra's light curve (top most panel) and is absent from the light curves of the other stars, suggesting this flare is unique to EK Dra and is not the result of local observing conditions.}
    \label{fig:ek_dra_im_and_flare}
\end{figure*}

The stability of Flarescope's light curves and, consequently, the ease with which we can identify flares are highly dependent on the weather.
The effects of poor observing conditions manifested as scatter up to $\sim10\%$ of the total stellar flux, especially for observations taking place during the winter season (modified Julian \added{dates} $\approx60280$--60390), which is a time of high precipitation at Palomar\footnote{Weather data at Palomar are archived by the \href{https://wrcc.dri.edu/cgi-bin/cliMAIN.pl?ca6657}{Western Regional Climate Center}.}.
Looking for flares in all 500 hours of Flarescope light curves revealed a few marginal ($<4\sigma$) flare candidates and one very-likely flare ($\approx26\sigma$ detection at flare peak). 
This flare, shown in Figures \ref{fig:ek_dra_im_and_flare} and \ref{fig:hi_res_flare}, occurred on EK~Dra and is the focus for the rest of this paper.

\subsection{Flare Validation}\label{sec:flare_validation}
The candidate flare from EK Dra peaked at UTC 2024-05-24, 06:35:40. 
The light curve for EK Dra (top right panel of Figure \ref{fig:ek_dra_im_and_flare}) around the time of the flare ($\approx 60\,$min into the start of the observation) follows the typical flare profile of a steep rise and exponential decay over $\approx20\,$min.
However, the observation stopped soon after the decay of the flare due to dangerous observing conditions---in this case, high humidity that suggested imminent rain.
An immediate question is then whether the \added{apparent }flare profile is a weather-based effect.
Looking directly at the 90\,s-integrated images themselves (including the image in the left panel of Figure \ref{fig:ek_dra_im_and_flare}), there is no apparent cloud coverage or change thereof. 
We should also note that EK~Dra is the brightest star in its field by at least an order of magnitude in \added{aperture-integrated} flux---for a local observing factor to increase its flux so substantially, the corresponding signal in the light curves of the other stars in the field would be an order-of-magnitude increase in their fluxes.  
Such a signal is not observed at any point in any of the light curves of the other stars in the field  (see the bottom four light curves of Figure \ref{fig:ek_dra_im_and_flare}) and so the signal is likely unrelated to weather conditions.

Alternatively, a satellite or plane could affect only EK Dra's derived flux. 
However, the duration of the flare candidate cannot be explained by \added{such} effects as \added{a plane or satellite} would pass through Flarescope's field of view (FOV, $\approx12\arcmin\times12\arcmin$) over the course of a few seconds, not minutes.
To be certain that the flare profile is not the effect of a satellite, we checked single-integration images (5\,s exposures).
We indeed find no streaks indicative of a satellite passing through the field at the time of the apparent flare. 
The distinctive shape and duration of the flare candidate in addition to the lack of an analogous signal on the other stars make us reasonably confident this is a genuine flare from EK~Dra.

\subsection{Flare Characterization}\label{sec:flare_char}
\begin{figure}
    \centering
    \includegraphics[width=\linewidth]{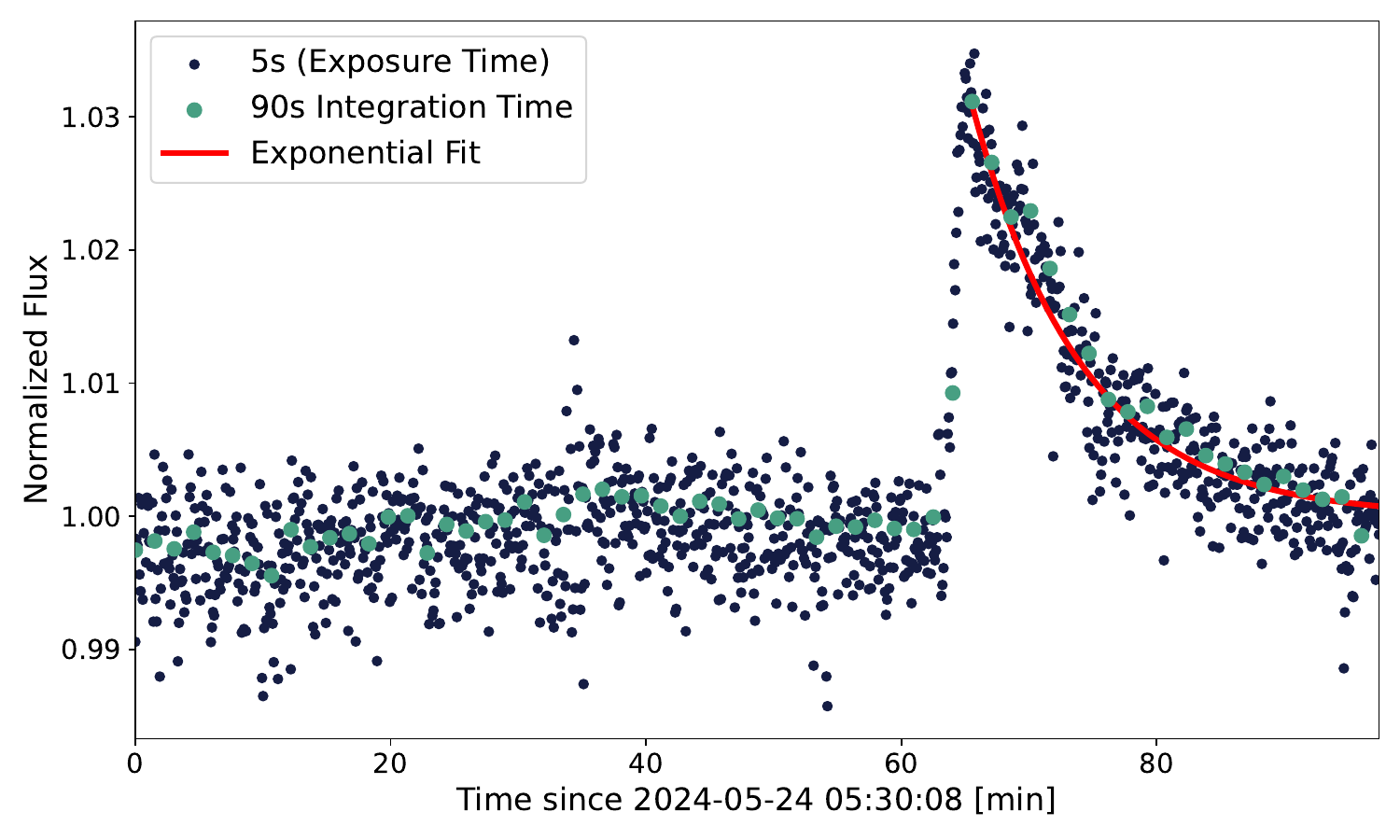}
    \caption{The EK Dra light curve at the native 5\,s integration time (dark blue) and the light curve at 90\,s integration time (teal). The best-fit exponential profile is also shown in red.}
    \label{fig:hi_res_flare}
\end{figure}

Although the large flare has some complex structure, its overall profile is dominated by the primary flare profile which is fit well by an exponential, shown in red in Figure \ref{fig:hi_res_flare}.
The best fit for this profile, estimated using \texttt{curve\_fit} from \texttt{scipy}'s \texttt{optimize} module \citep{Scipy2020}, has an amplitude of \added{0.031} relative to the median flux and an e-folding time, $\tau_{\text{decay}}$, of 8.59\,min. 
We use this profile \added{in place of} the light curve to estimate the bolometric flare energy, $E_{\text{flare}}$, from the normalized flare flux measured in the band, $F_{\text{band}}$.
This is done using the following equation:
\begin{equation}\label{eq:flare_flux}
    E_{\text{flare}} = F_{\text{band}}4\pi R_{\star}^2 \sigma_{\text{SB}} T_{\text{fl}}^4 \frac{\int_{g'}B(\lambda,T_{\text{q}})\text{d}\lambda}{\int_{g'} B(\lambda,T_{\text{fl}})\text{d}\lambda}\,t_{\text{int}}
\end{equation}
where $t_{\text{int}}$ is the integration time, $R_{\star}$ is the stellar radius,  $\sigma_{\text{SB}}$ is the Stefan-Boltzmann constant, $T_{\text{fl}}$ is the flare temperature, $T_{\text{q}}$ is the quiescent photosphere temperature, and $B(\lambda, T)$ is the Planck function for a blackbody over a range of wavelengths $\lambda$.

For this estimate, we use the exposure time which we used for the fit ($t_{\text{int}}=5\,$s), the quantities for EK~Dra reported in Table \ref{tab:stellar_info} for the radius and quiescent temperature, Sloan $g'$ wavelength range of 400--550\,nm, \added{and a constant flare temperature of 10,000\,K}.
\added{This flare temperature is standard for estimating energies from solar-type stars \citep[including EK~Dra specifically; ][]{Namekata2022}, although we note that it could very well be an underestimate of the true flare temperature \citep{Berger2023}.}
These quantities lead to a bolometric flare energy $E_{\text{flare}}\approx3.95\times10^{34}\,$erg; this may vary by a factor of a few depending on the actual flare temperature \added{and its }evolution.
The flare properties are summarized in Table~\ref{tab:flare_info}.

\begin{deluxetable}{l c}[h!!]
\tablecaption{EK~Dra flare properties assuming a flare temperature of 10,000\,K.}\label{tab:flare_info}
    \tablehead{
    \colhead{Property} & \colhead{Value}}
    \startdata
    {Flare Start} & 2024-05-24 06:33:24.5\\
    {Peak Time} & 2024-05-24 06:35:35.5\\
    {Peak Amplitude} & \added{$0.031$} \\
    $\tau_{\text{decay}}$ & 8.59\,min \\
    {$E_{\text{flare}}$} & $3.95\times10^{34}$\,erg\\
    $L_{\text{peak,bol}}$ & $7.83\times10^{31}\,\text{erg}\,\text{s}^{-1}$
     \enddata
\end{deluxetable}

 The energy we calculate is consistent with superflares observed from EK Dra, including those associated with massive prominence eruptions \citep{Namekata2022Nature, NamekataSuperflare2022, Namekata2024}.
 It is also worth noting from these works that a flare $>10^{34}\,$erg is expected to occur on EK~Dra about once every seven days---this is consistent with Flarescope only detecting one flare of this scale in the 189 hours it spent on EK~Dra. 
 The question, then, is whether this flare has associated radio emission indicative of particle flux.

\begin{figure}
    \centering
    \includegraphics[width=\linewidth]{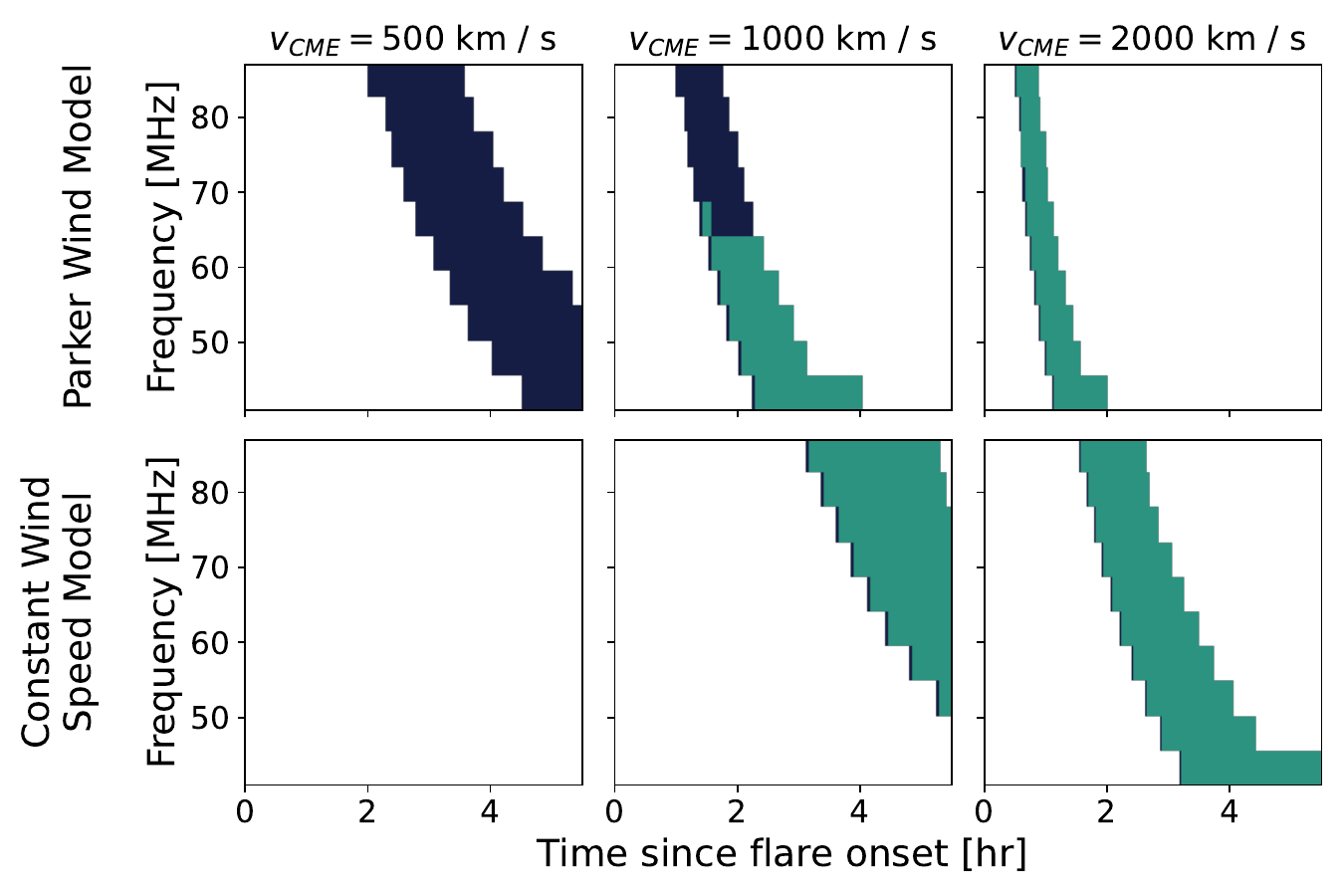}
    \caption{Example type~II dynamic spectra in the OVRO-LWA band for various CME speeds assuming that the CME expands linearly with distance. We show the shape of type~II bursts for times only within 5.5\,hr after the onset of the flare as it is the time frame for which we have radio data. Dark blue indicates time ranges where the CME would be exciting the correct frequencies but the speed is too low to drive a super-Alfv\'enic shock; teal indicates the frequencies and times that the CME can drive a shock.  Note that in the constant-wind-speed model, $v_{\text{CME}}=500\,$km\,s$^{-1}$  is too slow to reach frequencies in the OVRO-LWA band in the available time range (bottom left plot).}
    \label{fig:type_ii_shape}
\end{figure}

\begin{figure}
    \centering
          \includegraphics[width=0.8\linewidth]{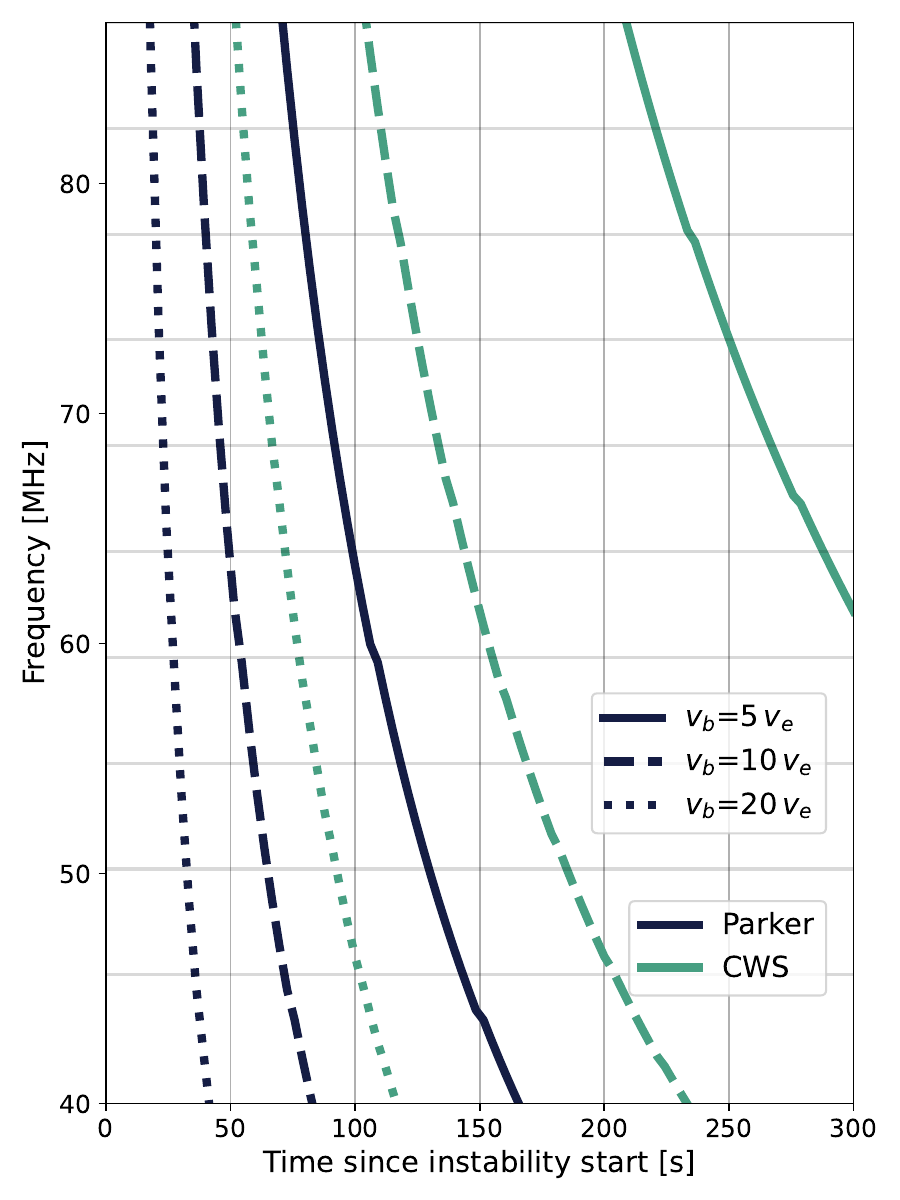}
    \caption {Example type~III burst shapes for beam speeds\added{, $v_\text{b}$,} at various factors of the thermal electron velocity $v_\text{e}=0.041\,c$ assuming the beam travels purely radially. We only demonstrate the contour of the burst edge and do not show the duration of the burst at a given frequency.}
    \label{fig:burst_shape_prediction}
\end{figure}

 Before evaluating the radio data, it is worth noting that the corona of EK~Dra is thought to be much hotter and denser than the solar corona \citep{Gudel1995, Ayres2015}---the properties of associated bursts may be substantially different from solar bursts.
 To illustrate how this will influence our search for bursts from EK~Dra, we consider two radial, isothermal wind models: one is a profile that satisfies the hydrodynamic conditions described by \citet{Parker1965} and the other profile simply assumes a constant wind speed (CWS) and dipolar magnetic field. 
 We take the base coronal electron number density to be $n_0 = 4\times10^{10}\,\text{cm}^{-3}$ and the coronal temperature to be $T_{\text{cor}}=10^7\,$K based on values derived in \citet{Gudel1995}.
 We take the \added{surface} field strength to be $1000\,$G, which is representative of values estimated in \citep{Kochukhov2020}. 
 We find that \added{a CME or electron beam may need to travel $\gtrsim10\,R_\star$ to excite plasma frequencies in the OVRO-LWA band. Thus,} it may take \added{several} hours for a super-Alfv\'enic shock to reach the OVRO-LWA observing band (see Figure~\ref{fig:type_ii_shape}). 
 For similar reasons, type~III bursts may have relatively shallow drifts, as illustrated in Figure~\ref{fig:burst_shape_prediction}.
 This influences both the time range of data that we evaluate as well as the de-dispersion parameters we use for the burst search.
 More details on how we approach the wind modeling are provided in Appendix~\ref{app:wind_profiles}.

\section{OVRO-LWA Data Reduction}\label{sec:ovro_lwa_data_reduction}
Due to a data recorder failure, we do not have the high-time-resolution beamformed data for this observation;
we only focus on the cross-correlated data here. 
The ultimate goal is to derive a coarse dynamic spectrum for EK~Dra from the images produced from cross-correlated data.
We focus on the frequency range 41--87\,MHz \added{as it} had reasonable calibration solutions\added{. We} divide the data into 10, 4.6\,MHz-wide subbands for imaging.
As discussed in the previous section, it may take hours for a CME to reach the densities associated with plasma frequencies in the OVRO-LWA observing band. 
Alternatively, a type~III burst may happen essentially any time relative to the flare, but is likely to be brightest near the time of the flare.
Given such broad constraints for the arrival time of a burst to the OVRO-LWA band, we evaluate all data available for the night of the flare: UTC hours 05--12 (hours 22--05 local time).

The data were calibrated in three stages: delay calibration, bandpass calibration, and per-snapshot imaging.
The pipeline for this calibration will be discussed in detail in forthcoming papers, but we summarize the steps here as they were adapted for this dataset.

\paragraph{Delay calibration}{For each subband in the 41--87\,MHz range, the pipeline identifies the measurement set (MS) integration with a local-sidereal time (LST) closest to Cygnus~A's (Cyg~A) meridian transit; these MSs are concatenated into a single MS that represents the full 41--87\,MHz range and are re-gridded using \texttt{CASA}'s \texttt{mstransform} function so that the delay solver will see the full instantaneous bandwidth \citep{CASA2022}. We flagged bad antennas based on the same stability standards described in \citetalias{Davis2025}--- this led to 46 of the 352 antennas being flagged alongside additional baseline flagging from \texttt{AOFlagger} run with default parameters \citep{AOFlagger2012}. A beam-weighted model of Cyg~A is Fourier-transformed into the data column of the MS and \texttt{gaincal} is run in the K-delay mode across baselines 10--125$\,\lambda$ to solve for the antenna delays. The quality of the delay calibration is confirmed by eye.}
\paragraph{Bandpass calibration}{Bandpass calibrations are derived from MS files spanning four minutes nearest Cyg~A's transit. A sky model is made which includes all bright sources above the horizon; for this observation, the dominant bright sources are Cyg~A and Cas~A. The phase center is shifted to zenith of the middle snapshot so that all scans share the same direction-independent gains. After applying flags, \texttt{CASA}'s \texttt{bandpass} task solves for amplitude- and phase-versus-frequency using the delay table derived in the previous step.  These bandpass and delay solutions are applied to all integration times using \texttt{applycal}}.
\paragraph{Imaging}{Because there were no exceptionally-bright sources near EK~Dra, we did not need to aggressively clean the data; Cyg and Cassiopeia~A (Cas~A) being low on the horizon posed the biggest limitation to producing high-quality data.
We did a maximum of five major iterations of peeling with \texttt{ttcal}'s \texttt{zest} functionality in an attempt to remove the effects of sidelobes from these sources without leading to the peeling algorithm diverging due to the sources' low elevations \citep{eastwood2017}.
Stokes I and V images were produced for each integration time in each subband using 20,000 cleaning iterations and using an inner tukey taper of 30 wavelengths, a Briggs weighting of 0, horizon mask of $10^\circ$, and a pixel resolution of $1.875\arcmin$ as inputs for \texttt{wsclean} \citep{wsclean2014}.}

\begin{figure}
    \centering
    \includegraphics[width=\linewidth]{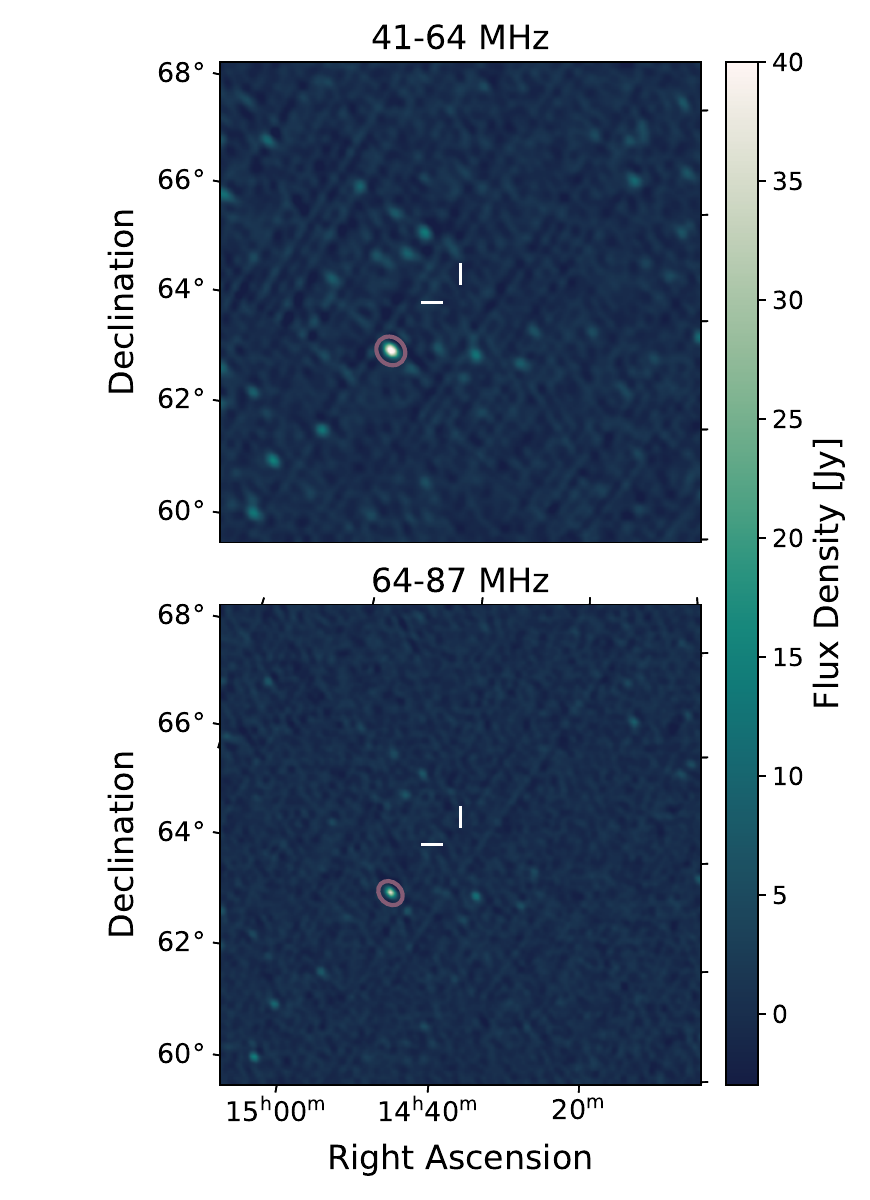}
    \caption{Example snapshot images integrated between 41--64\,MHz (top panel) and 64--87\,MHz (bottom panel), cropped around an $8^\circ\times8^\circ$ region centered on EK~Dra. The position of 3C~305---the only consistently-detected source in this observation---is indicated by an ellipse and the expected position of EK~Dra is indicated by a white crosshair. }
    \label{fig:ovro-lwa-fov}
\end{figure}

Some of the subbands were missing timestamps, possibly due to the data recorder failing on some intervals, that led to the loss of a total of 291 integrations, or about 1.2\% of the data.
The amount of flagged or unavailable data increases to $6\%$ after flagging data during the hourly beacon signals from the Los Angeles Department of Water and Power (LADWP).
After the entire sky is imaged, we crop the images to $256\times256$\,pix $\approx8^\circ\times8^\circ$ centered on EK~Dra.
Example snapshot images for the 41--64\,MHz and 64--87\,MHz ranges are shown in Figure~\ref{fig:ovro-lwa-fov}. 

\subsection{Position Estimates and flux density extraction}
\begin{figure*}
    \centering
    \includegraphics[width=0.8\linewidth]{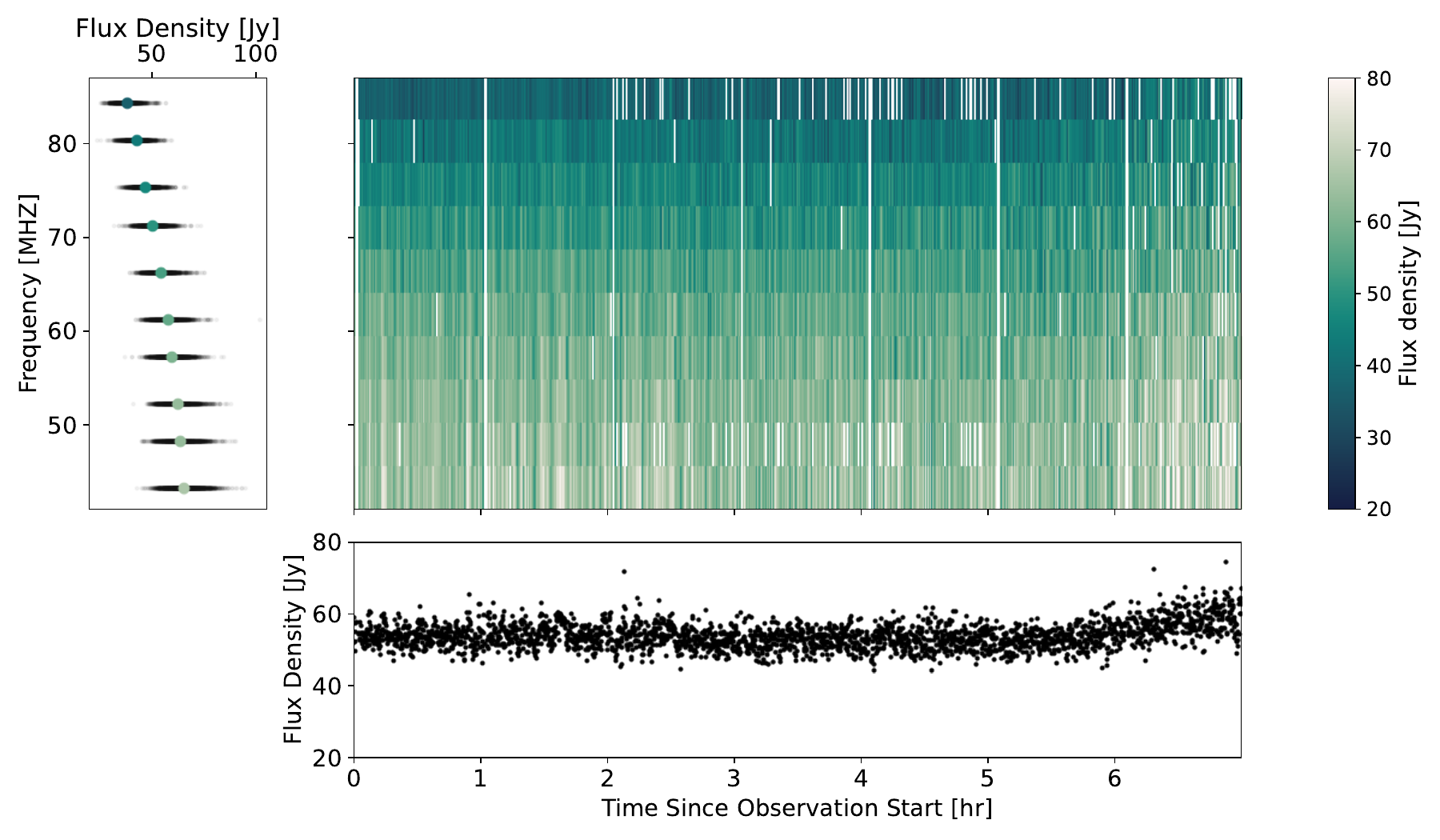}
    \caption{3C~305's dynamic spectrum (top right), per-integration (black points) and time-averaged (colored points) spectra (top left), and frequency-averaged light curve (bottom plot) derived from the cross-correlated data. The periodically flagged data are due to the LADWP beacon pulses. The flux density has been corrected for the elevation of the source throughout the observation; the apparent increase in the flux density at the end of the observation may be due to a poor correction at low elevations ($\approx30^\circ$). The stability in the extracted flux density is one indicator that the ionosphere was relatively well-behaved during this observation.}
    \label{fig:bright_source_dynspec}
\end{figure*}

Reliably extracting a dynamic spectrum from a star's location requires we account for the ionosphere's effects on the apparent position of sources in the sky.
Because these effects depend on time, position, and frequency, we rely on the motion of other sources in the sky to predict EK~Dra's apparent position.
The radio sky in the vicinity of EK~Dra is relatively sparse; there is only one source (3C~305, indicated by an ellipse in Figure~\ref{fig:ovro-lwa-fov}) that is consistently detected in each of the 10 subbands with which we may use to estimate EK~Dra's apparent position on the sky.
While this might normally be a problem, the ionosphere during this observation was remarkably well behaved; the median source position variability it caused was on the order of $10\arcsec$, well within the size of the typical beam of $\approx15'\times12'$.
This stability is corroborated by the dynamic spectrum we extracted for 3C~305, shown in Figure \ref{fig:bright_source_dynspec}. 
This dynamic spectrum shows flux-density variance of a few Janskys, consistent with expectations for a source this bright.
We are thus comfortable using the position of the singular source to estimate the position of EK~Dra on the sky.
The extraction of the flux density at the location of EK~Dra in each sub-band then follows the same steps as the extraction procedure presented in \citetalias{Davis2025}; the resulting dynamic spectrum and light curve are shown in Figure \ref{fig:ek_dra_dyn_spec}.

\section{Radio Results}\label{sec:ovro_lwa_results}
\begin{figure*}
    \centering
    \includegraphics[width=0.8\linewidth]{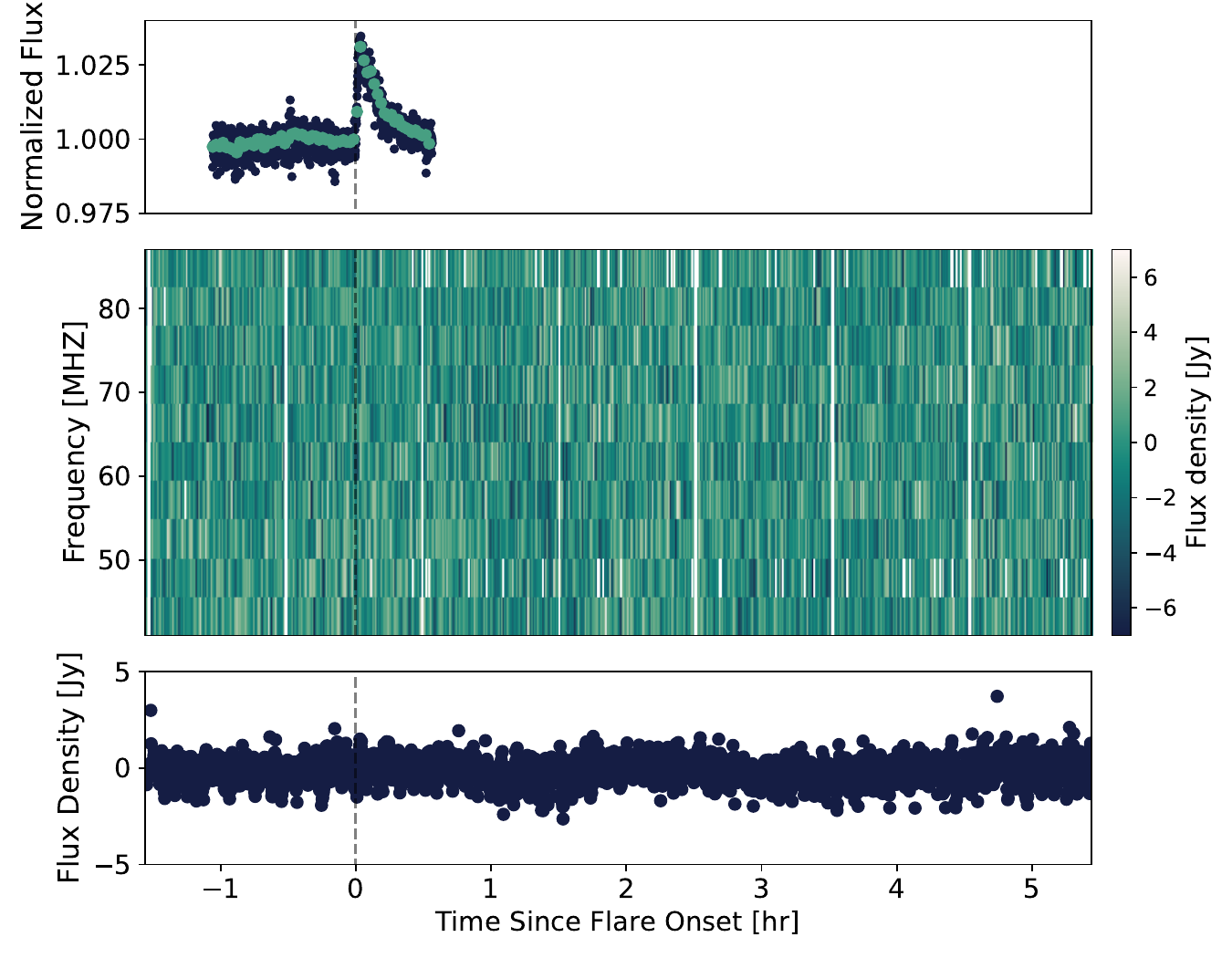}
    \caption{The dynamic spectrum for EK~Dra (middle panel) along with the light curve extracted from averaging over the entire dynamic spectrum (bottom-most panel). The Flarescope light curve is included in the top-most panel to illustrate when the flare occurred relative to the rest of the observation. }
    \label{fig:ek_dra_dyn_spec}
\end{figure*}

At the 4.6\,MHz bandwidth and 10\,s integration time, there is no apparent type~II or III signal in EK~Dra's dynamic spectrum.
It is very plausible that the noise---on average 1.73\,Jy in the highest (82--87\,MHz) subband and 2.29\,Jy in the lowest (41--46\,MHz) subband---is too high to observe the burst in any given subband.
After averaging across 41--87\,MHz to produce the light curve (bottom most panel of Figure \ref{fig:ek_dra_dyn_spec}), there are a few integrations scattered about the time range that suggest a possible signal; checking these integrations by eye in the image plane make it clear there was anomalously-high noise during those integrations and there is no signal from EK~Dra.
The question then is if de-dispersing before integrating across frequency reveals a signal.

\subsection{Type~III Burst Search}
If EK~Dra's corona were like the Sun's, then we might expect that a type~III burst \added{would drift too fast and be too short to} benefit from a de-dispersion search in cross-correlated data. 
However, the density profiles we estimate for the wind predict much shallower drifts, demonstrated in Figure \ref{fig:burst_shape_prediction}.
Additionally, the velocity dispersion of the electron beam exciter elongates the beam as a function of distance---the further a beam travels, the longer the excitation lasts at the corresponding frequency \citep{ReidRatcliffe2014}.
Because the beam may need to travel $\approx10\times$ further than a beam \added{on} the Sun \added{to excite the OVRO-LWA frequencies}, we might expect that \added{a} burst \added{from EK~Dra }lasts $\approx10\times$ longer than a solar type~III burst.
If this is the case, bursts would last on the order of minutes instead of $\approx10\,$s, enabling us to bin in time without the risk of losing signal.

\begin{figure}
    \centering
    \includegraphics[width=\linewidth]{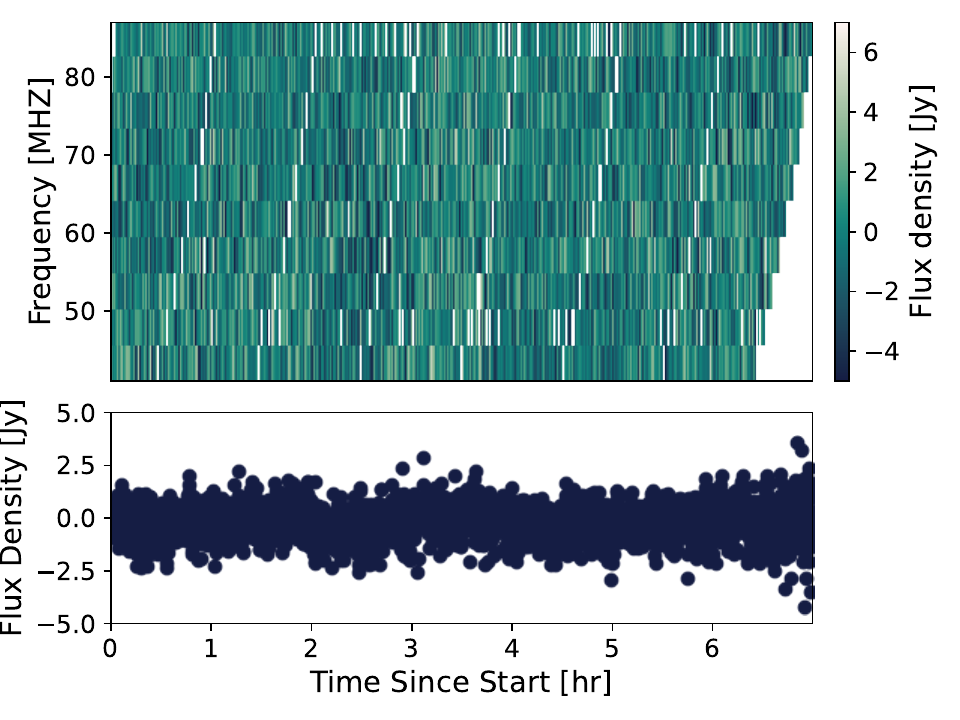}
    \caption{An extreme example of a ``de-dispersed" EK~Dra dynamic spectrum (top panel) and the resulting light curve from averaging over all frequencies (bottom panel) for $a=-3.7$ and $\alpha=1.1$.}
    \label{fig:dedisp_dyn_spec}
\end{figure}

Before starting the search, we mask the data at the timestamps and 20\,s around the times of the DWP signal, as identified in the dynamic spectrum of 3C~305.
We estimate the delay of a given frequency, $\nu_i$, relative to the top of the band, $\nu_{87}$, using the expression provided by \citet{AlvarezHaddock1973DriftRate}: 
\begin{eqnarray}
    d\nu/dt = -10^{a}\nu^\alpha \,\text{MHz\;s}^{-1} \rightarrow \nonumber\\
    \Delta t = -10^{-a}(1-\alpha)^{-1}(\nu_{87}^{(1-\alpha)} - \nu_i^{(1-\alpha)})\,\text{s},
\end{eqnarray}
Because we expect the drift rate to be slower than for the Sun, we prioritize evaluating $a$ and $\alpha$ values smaller than the solar estimate ($a=-2$, $\alpha=1.84$); we investigate $-5<a<-1$ and $1<\alpha<3$.
After de-dispersing for each $a$ and $\alpha$ value, we average over frequency.
An example of this de-dispersion and frequency integration is shown in Figure \ref{fig:dedisp_dyn_spec}. 
This is done at the native integration time, after binning up to 1\,min (six integrations), and after binning up to $2\,$min.
This search did not reveal any signal that could not be explained by a contaminating, poor-quality integration; there is no detectable type~III burst at the $5\sigma=3.2\,\text{Jy}$ threshold \added{at the 10\,s integration time}.
\added{This upper limit corresponds to the equivalent of a $1.6\times10^{10}\,$SFU burst from the Sun, or about 4--5 orders of magnitude brighter than the brightest solar type~III bursts at these frequencies \citep{Saint-Hilaire2013, Sasikumar2022}.}

\subsection{Type II Burst Search}
\added{Because type II burst} drifts \added{rates depend} on both the \added{propagation} environment and the speed and geometry of the CME\added{, the parameter space for de-dispersion is unconstrained compared to that of type~III bursts.}
\added{Because of this, and because a type~II search can benefit from much higher integration times, we do not attempt a de-dispersion search for a type~II burst.}
\added{Instead, we} bin in time to 2, 5, and 10\,min time resolutions and shift the binnings by 1\,min intervals to account for possible signal loss due to excessive background inclusion.
\added{We} perform various frequency \added{averagings} for each time resolution, including averaging over two, three, five and all ten subbands, corresponding 9.2\,MHz, 13.8\,MHz, 23\,MHz, and 46\,MHz, respectively.
We look for a signature of a type~II burst in these extra-coarse dynamic spectra and light curve. Again, there is no emission that can be attributed to the star\added{. For the 10\,min light curve, this corresponds to a $5\sigma$ detection threshold of 1.75\,Jy, or the equivalent of a $8.8\times10^9\,$SFU burst from the Sun}. 
EK~Dra produced no detectable radio emission in the 5.5\,hr following its flare that would indicate material escaped into the interplanetary medium.

\section{Discussion}\label{sec:discussion}

Any solar-scale radio burst produced at the distance of EK~Dra would not be detectable by the OVRO-LWA.
However, the prominence masses that have been reported for EK~Dra are orders of magnitude greater than solar CMEs\added{. A}s previously iterated, the environment of the star is \added{also} substantially different from the Sun---it would not be unreasonable to expect exceptional radio bursts.  
The question then is whether the lack of a type~II or III detection at the OVRO-LWA's sensitivity is a function of the stellar properties, the CME/plasma beam properties, or related to the fundamental mechanics of type~II and III bursts.
Here, we present the different conditions that may lead to us not detecting a type~II or III burst associated with \added{the} flare \added{reported here}.

\subsection{Type III Non-detection}
It is worth emphasizing here that type~III bursts are the most common burst on the Sun---essentially all solar flares produce them and they even occur in the absence of X-ray flares. 
The reconnection process intrinsically opens magnetic field lines---there are almost certainly beams of electrons being produced during this EK~Dra flare\added{. The} simultaneity of the Flarescope and OVRO-LWA observation allows us to investigate observational and physical limitations in a way that is not normally afforded to stellar analyses below 100\,MHz.
We may not see a type~III burst if 1.) the beam velocity is insufficient to develop the bump-on-tail distribution, 2.) the instability growth rate is insufficient to facilitate the effective conversion to Langmuir waves, 3.) the magnetic field line leading the beam does not extend out to the distance where the density is conducive to being observed by the OVRO-LWA, or 4.) the burst brightness is limited by the saturation limit of the plasma emission mechanism.
We consider each of these possibilities here.

\subsubsection{Beam velocity and instability development}
The bump-on-tail instability requires that the velocity of the propagating beam $v_{\text{b}}$ is higher than the background thermal speed of electrons $v_{\text{e}} = \sqrt{k_{\text{B}}\,T\,m_{\text{e}^-}^{-1}}$ where $k_{\text{B}}$ is the Boltzmann constant and $m_{\text{e}^-}$ is the electron mass.
EK~Dra's coronal temperatures of 2--10\,MK are $\approx1$--$10\times$ the quiescent solar temperature and require that $v_\text{b}\gtrsim0.04\,c$.
In actuality, we see for solar bursts that $v_\text{b}/v_\text{e}\approx10$--50 \citep{Dulk1985}.
This may then imply that we need a beam with velocity $v_\text{b}\gg0.4\,c$ to develop the anisotropy.
This exceptionally high speed requirement is relaxed if we assume the temperature at 10s of stellar radii is much lower than the coronal temperature; this assumption is consistent with what is observed for the Sun and that has been estimated for T Tauri stars \citep{JohnsKrull2007TWHyd}.
It is thus likely that a beam reaching 10s of stellar radii would be capable of reaching a speed sufficient to develop the instability needed to excite Langmuir waves.

The development of the instability also requires that its growth rate is much larger than the diffusion rate.
However, both of these rates depend on quantities that are not reasonable to speculate for EK~Dra, especially as the beam evolves throughout the stellar atmosphere; we will assume that if the beam moves fast enough to develop the anisotropy then it is also capable of facilitating the wave growth. 

\subsubsection{Beam directionality}\label{sec:typeiii.beamdirection}
Although solar type~III bursts are most commonly observed as consistently propagating to lower frequencies, there are a few records of type~III bursts (and, presumably, their driving beam) changing direction \citep{MaxwellSwarup1958TypeU}.
Such bursts---called type U or J bursts due to the shape that the reversal produces in dynamic spectra---are associated with fields confined in the solar corona.
Given the extreme distances that EK~Dra's beam needs to travel to excite the necessary frequencies for the observation, it is not unreasonable that an opened field line may kink in such a way that the beam does not reach the relevant densities for the OVRO-LWA.
Observing at higher frequencies may then reveal type~III bursts, although the low altitudes and corresponding high temperatures probed by such frequencies may then reintroduce the velocity distribution problem mentioned previously.

Discussion would normally also involve the domination of the magnetic field and thus magnetic emission processes at such low altitudes.
However, even when assuming a base magnetic field strength of $\approx1$\,kG like \citet{Kochukhov2020} suggest EK~Dra produces, the plasma frequency is larger than the cyclotron frequency for distances \added{$\gtrsim0.5\,R_\star$} above the stellar surface in the CWS model or \added{$\gtrsim0.9R_\star$} in the Parker wind model; radio emission from EK~Dra that is detected \added{$\lesssim1000\,$}MHz is not likely to be cyclotron emission.

\subsubsection{Brightness limitations}{
Finally, it may be that any burst EK~Dra could produce simply is not bright enough to be observed by the OVRO-LWA.
To evaluate this possibility, let us consider the brightness temperature and source size to estimate what the flux density would be. 
The brightness temperature of plasma emission is limited by the energy density of Langmuir waves and their associated temperature, which in turn is limited by the thermal electron temperature.
For a thermal plasma with temperature $10^6\,$K---which we will assume is representative of the temperature of EK~Dra's atmosphere where the burst originates even though the corona may be much hotter---\citet{Melrose1989} estimates the limiting brightness temperature of plasma emission to be $\approx10^{17-18}\,$K.
The median source size of type~III bursts at 150\,MHz is $5.3\arcmin$ when observed at a distance of $\sim1\,$AU \citep{Saint-Hilaire2013}; we might expect it to be $\gtrsim15\arcmin$ in the OVRO-LWA band.
Thus, if EK~Dra's type~III burst is produced at the plasma emission limit for brightness temperature, the burst would have a flux density of $\approx30\,$Jy.
\citet{Vedantham2020} reduces this quantity by a factor of 100 based on solid angle broadening; this would reduce our estimate to $0.3\,$Jy, below the sensitivity threshold of the OVRO-LWA.

This might suggest that based on physical limitations of the plasma emission process, type~III bursts from the distance of EK~Dra would be undetectable by the OVRO-LWA.
That said, the propagation requirements for EK~Dra may again work in our favor; it might be reasonable to expect that the type~III burst source size is much larger due to how much further the beam would need to travel, both addressing the flux density limit as well as the saturation extremity. 
This is to say that it should not be unreasonable for detectable emission to be produced within the physical limits of plasma emission, but that a detectable burst may not have occurred during this observation.
}
\subsection{Type~II Non-detection}
Although type~II bursts are also the result of plasma emission, the conditions of their production are different from---and much less well-constrained than---type~III bursts.
Depending on the wind model assumed, a \added{shock front} may need to travel \added{$8$--$21R_\star$} to excite frequencies in the OVRO-LWA band (see Appendix~\ref{app:wind_profiles} and Figure~\ref{fig:type_ii_shape}).
Assuming \added{a }CME \added{only occurs after} the flare, then it would \added{have needed} to drive a shock at these distances within the 5.5\,hr after the flare for \added{us to have observed it with the OVRO-LWA}. 
If \added{the CME} were capable of doing this, then we \added{could} put limits on an associated type~II luminosity and evaluate it relative to the scaling relation introduced by \citet{Mohan2024_scaling_relation}, which shows a relationship between type~II luminosity, X-ray flare luminosity, and CME speed. 
\added{We explore these speed requirements and flare-CME relationship here. We also consider the potential effects of CME propagation geometry on burst detectability.}

\subsubsection{CME Velocity: Band arrival time and shock development}
\begin{figure}
    \centering
    \includegraphics[width=0.95\linewidth]{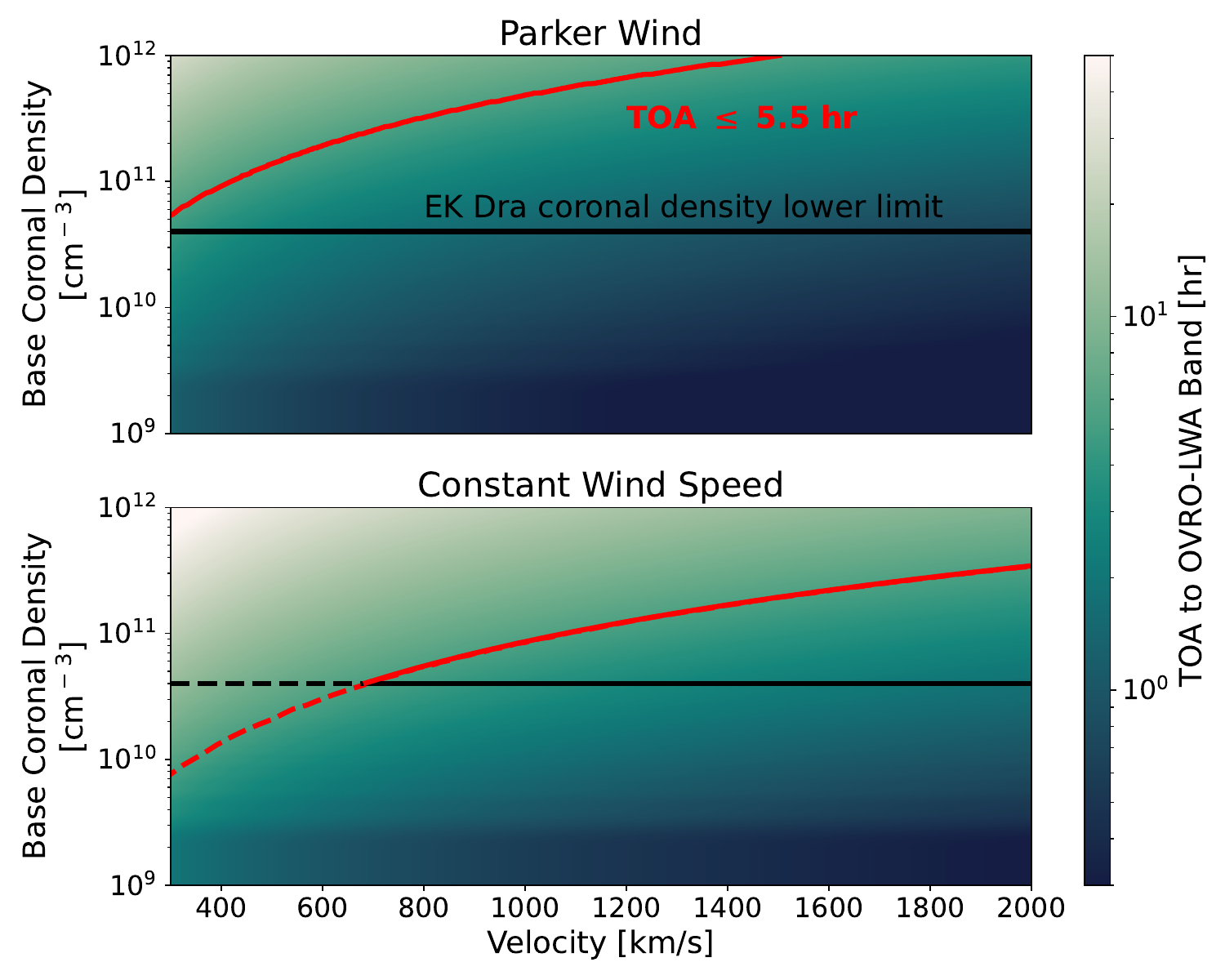}
    \caption{Time-of-arrival (TOA) estimates for the CME to excite plasma frequencies in the OVRO-LWA band for the Parker wind (top panel) and CWS (bottom) models. The red line indicates the parameters that allow the CME to reach the band within 5.5\,hr and the black line indicates the lower limit on the base coronal density from \citet{Gudel1995}.}
    \label{fig:toa}
\end{figure}

Assuming that \added{the} flare \added{presented in this work} has an associated CME, a simple explanation is that the CME did not travel fast enough to 1.) reach a distance within EK~Dra's wind where the density is conducive to producing plasma emission \added{in} the OVRO-LWA band or 2.) drive a super-Alfv\'enic shock. 
To visualize the arrival-time constraint, we consider base coronal densities $10^{9}<n_\text{b}<10^{12}$, the lower limit of which is representative of the solar corona and the upper limit of which is an order of magnitude lower than the high-end of stellar coronal densities \citep{Gudel2004}. 
We also consider $300<v_\text{CME}<2000\,$km/s; this parameter space is represented in Figure \ref{fig:toa}.
Taking the lower limit on the estimate for EK~Dra's base density ($4\times10^{10}\,\text{cm}^{-3}$) and ignoring that CMEs expand as they propagate, we find that the speed would need to be at least \added{680}\,km\,s$^{-1}$ in the CWS model to reach the OVRO-LWA observing band; the speed required in the Parker wind model is  $<300$\,km\,s$^{-1}$.

While the Parker wind \added{CME-}speed requirements should be reasonably achievable, the speeds required in the CWS model are \added{at the upper end of} line-of-sight (LOS) speeds that have been reported for EK~Dra \citep{Namekata2025V889Her}.
However, the LOS speeds are likely underestimates by virtue of the nature of the measurement; assuming equipartition of flare and CME kinetic energy, the CME could be moving anywhere 850--5150\,km/s when considering the mass range of prominences for EK~Dra ($\sim10^{17-19}\,$g).
Although strong overlying fields in principle could slow a CME, reports of velocities for CME candidates on other highly active stars have achieved these speeds \citep[see, e.g., ][ for a summary]{Moschou2019}. 
This is to say that it is not unreasonable to expect a CME to move fast enough to show up in the OVRO-LWA's band, but also a ``low" speed could very easily explain why we did not see a type~II signal.

Even if the CME is fast enough to have a reasonable TOA, we may not observe a type~II burst if \added{the CME} was unable to drive a super-Alfv\'enic shock in the frame of the wind. 
This may be a limiting factor in the Parker wind model; the fast winds this model predicts require speeds \added{$\gtrsim960$}\,km\,s$^{-1}$ to produce a shock. 
Conversely, in the CWS model, any CME that would have arrived to the OVRO-LWA band in time would have been able to drive a shock.
While the timing is largely an observational constraint, the shock development is a physical constraint; this analysis would benefit greatly from more robust modeling of EK~Dra's corona and wind.

\subsubsection{Under-luminosity}
\begin{figure}
    \centering
    \includegraphics[width=\linewidth]{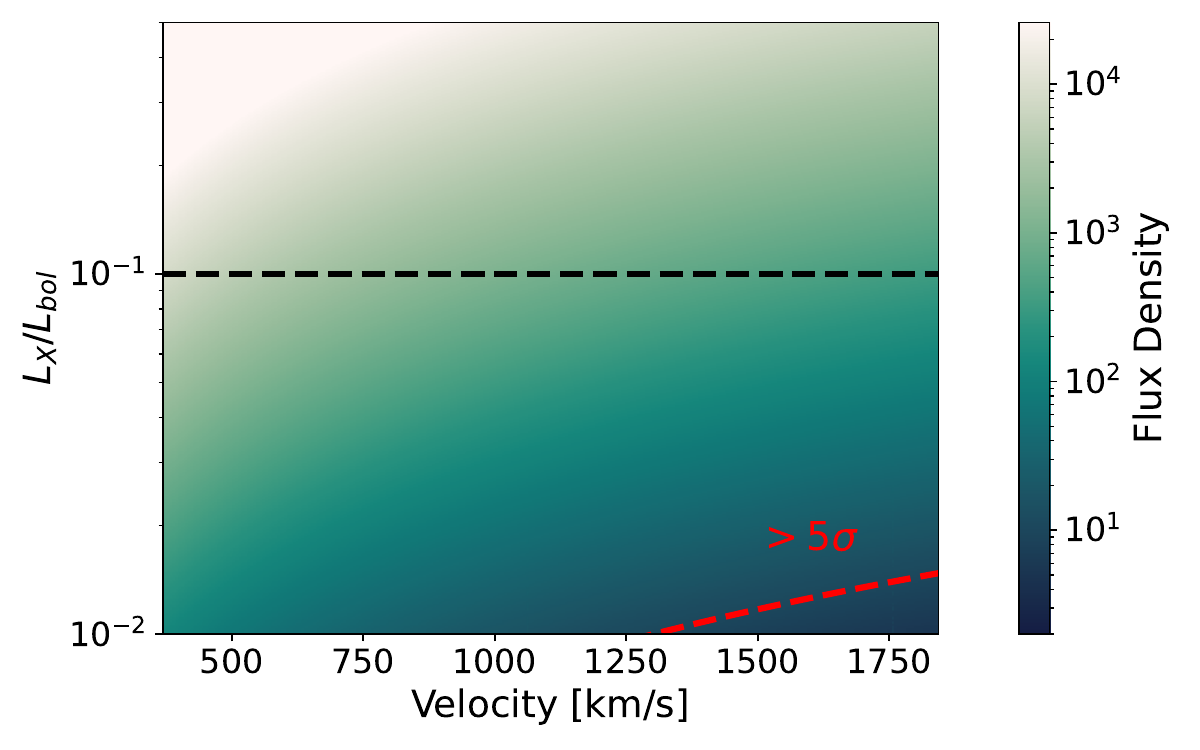}
    \caption{The anticipated flux density of EK~Dra, calculated from equation \ref{eq:type_ii_LR}. An X-ray luminosity that is $10\%$ of the bolometric flare luminosity is indicated by a black dashed line, as this is the typical partitioning of energy expected of flares. The red indicates flux densities above which a burst would have been a $5\sigma\approx10\,$Jy detection in any 4.6\,MHz-wide subband. This suggests any reasonable partitioning of the bolometric flare luminosity to the X-ray would have a corresponding type~II radio luminosity that is detectable by the OVRO-LWA, assuming EK~Dra follows the same scaling relation as the Sun. }
    \label{fig:type_ii_flux}
\end{figure}

If the CME was actually fast enough to satisfy both the TOA and shock conditions, then an alternative explanation is that the type~II burst simply was not bright enough to detect.
To explore this, we consider \citet{Mohan2024_scaling_relation}'s scaling relation between CME velocity, X-ray luminosity, and type~II radio luminosity.
This is given as:
\begin{equation}\label{eq:type_ii_LR}
    \log_{10}{(L_{\text{X}})} \approx 1.12\log_{10}(\sqrt{L_{\text{R}}\,v_{\text{CME}}^2}) + 9.5,
\end{equation}
where all quantities are in centimeter-gram-second units. 
We consider X-ray luminosities between 1--50\% of the peak bolometric flare luminosity and, because \citet{Mohan2024_scaling_relation}'s scaling relation comes from frequencies outside of the OVRO-LWA band and thus may be more luminous than in our band, we reduce the anticipated flux densities by a factor of 10 to produce the flux density estimates in Figure \ref{fig:type_ii_flux}.

Even with this reduction to anticipated flux density, the resulting type~II burst should have produced a $\gtrsim5\sigma$ signal so long as the X-ray luminosity was $\gtrsim2$\% of the bolometric luminosity---such an X-ray contribution is entirely reasonable given typical estimates that it makes up $\approx10\%$ of the bolometric flare energy \citep{OstenWolk2015}.
This is to say that, assuming the CME was fast enough and stellar bursts follow solar scaling laws, then a type~II burst should have been luminous enough to observe.
The latter caveat is, of course, a major limitation to our ability to determine detectability.
There has been one plausible stellar type~II burst, which peaked at $440\,$mJy from a distance of 40\,pc\added{, or the equivalent of a $\approx3\times10^9$\,SFU burst from the Sun} \citep{Callingham2025} in the LOFAR high band array (110--190\,MHz)\added{. However,} there was no flare observed contemporaneously to evaluate the relation from \citet{Mohan2024_scaling_relation}.
Extrapolating from solar to stellar quantities is even more dubious when considering the importance of shock formation on type~II production and the incredibly different conditions that EK~Dra presents for shock formation \added{compared to the Sun}.
Although it would be useful to use flux density estimates to put constraints on the CME velocity, the solar scaling law may, in fact, only be credible for the Sun.

\subsubsection{Confinement}
\added{We have so far been assuming} that there had actually been a CME to detect. 
However, the ever-present problem of stellar CME studies is being able to conclusively say that material escaped.
As an order-of-magnitude estimate, we may say that a CME would be confined if pressure from the overlying magnetic field is greater than the outward pressure of the CME.
We estimate the CME pressure as its kinetic energy, $E_{\text{kin}}$, divided by its volume; we estimate the flare loop volume to be $V\approx3.6\times10^{30}\,\text{cm}^3$ (see Appendix~\ref{app:flare_loop}) and we consider CME energies between 0.01--10 times the flare energy.

We take the magnetic pressure to be simply $B^2/8\pi$ where $B= B_0(R/R_\star)^{-3}$ and $B_0$ is the surface magnetic field strength. 
Zeeman-Doppler imaging and modeling reported in \citet{Namekata2024b} found maximal global field strengths of $\approx200\,$G.
At the distance of the top of the flare loop ($\approx0.4\,R_\star$ above the stellar surface from estimations in Appendix~\ref{app:flare_loop}) the magnetic field would be $\approx73\,$G.
The ratio of the CME kinetic pressure to the magnetic field pressure would then be $\approx0.3$--270 depending on the partitioning of CME and flare energy, where a ratio $<1$ implies the magnetic pressure dominates and thus inhibits CME escape.
We replicate these calculations using the field strength estimate of 1.4\,kG from \citet{Kochukhov2020} and get ratios of 0.006--6, implying, as we would expect, that confinement is much more likely \added{when a stronger field is present}; relevant quantities and their resulting pressure ratios are summarized in Table \ref{tab:cme_confinement_summary}.

\begin{deluxetable*}{cccccccc}
\tablecaption{A summary of the quantities used to estimate the CME pressure $P_\text{CME}$ and magnetic pressure $P_B$.}\label{tab:cme_confinement_summary}
\tablehead{
 \colhead{$L$} & \colhead{V} & \colhead{$E_\text{Kin}$} & \colhead{$P_\text{CME}$} & \colhead{$B_0$}  & \colhead{$B(r=0.4\,R_\star)$} & \colhead{$P_B$} & \colhead{$P_\text{CME}/P_B$}\\
 \colhead{$R_\star$} & \colhead{cm$^{3}$} & \colhead{erg} & \colhead{erg\,cm$^{-3}$} & \colhead{G} & \colhead{G}&\colhead{erg\,cm$^{-3}$}& \colhead{{}}}
        \startdata
         $0.4\,R_\star$& $7\times10^{30}$& $4\times10^{32}$ & 57& 200 & 73 & 211  & 0.3\\
         & & & & 1400 & 510 & 10,360  &  0.006\\
        {}& {}& $4\times10^{35}$& 57,140& 200& 73 &211  & 271 \\
        {}& {}& & & 1400& 510 &10,360 & 6
        \enddata
\end{deluxetable*}

This rough estimate suggests that there is a range of CME energies \added{associated with this flare} where \added{EK~Dra's }overlying magnetic fields can reasonably prevent the CME from escaping.
When considering additional complications introduced by, e.g., ejection geometry and fragmentation \citep{AlvaradoGomez2018, AlvaradoGomez2022}, it becomes easy to imagine that this could have been a scenario where there was a failed eruption---material would have never gone further than a couple stellar radii, let alone the tens of radii required in the CWS model.
In this case, it is worth again noting the exceptional energy of the flare, which was on the same scale or larger than flares from EK~Dra that have produced fast, massive prominences.
Although a lack of a CME signature at this flare energy does not necessarily mean a CME could \textit{never} occur at this energy, it could suggest, for instance, that mass loss through CMEs does not contribute nearly as much to overall mass loss rates as has been suggested \citep{OstenWolk2015, Cranmer2017CME}.
That said, the prediction for CME confinement presented here is far from rigorous; this discussion would benefit greatly from similar wind and CME modeling done in \citet{AlvaradoGomez2018} and \citet{AlvaradoGomez2022} when accounting for EK~Dra's exceptional coronal environment.

\added{\subsection{Propagation Geometry}
We have so far only considered the radial component of the wind in our modeling. 
However, we know for the Sun---and may assume for other solar-type stars---that the rotation of the Sun introduces a spiral structure to the wind. 
Impulsively-accelerated particles generally follow this path.
Because the material needs to propagate so far to excite a given frequency, the material may have followed the spiral behind the star by the time it reaches the relevant densities.

Here, we consider the simplest interpretation of a spiral wind where, in a time interval $\Delta t$, a wind parcel  travels a distance $v_w\Delta t$ and EK~Dra rotates an amount $2\pi\Delta t/P $ where $P=2.6\,$d is the rotation period of EK~Dra that we derive from \emph{Transiting Exoplanet Survey Satellite} (TESS) light curves \citep{Ricker2014}.
The resulting curvatures for the CWS and Parker wind models are shown in Figure~\ref{fig:spiral} along with an approximation for the physical extent that a CME or electron beam may have.
As with our demonstration of potential type~II burst shapes, we made this estimate by assuming that the exciter expands linearly with distance from its point of origin.
We take the point of origin to be the top of the $0.4\,R_\star$ loop that we calculate in Appendix~\ref{app:flare_loop} and that the initial extent of the beam is $0.1\,R_\star$.
 This is an incredibly simplified treatment of the complex dynamics at play. 
 That said, using solar quantities \citep[starting height of $0.1\,R_\odot$ and initial extent of $0.1\,R_\odot$,][]{Aschwanden2000, Gopalswamy2000} and an expansion factor of 0.5 replicates the $\sim0.5\,$au size of solar CMEs at $1\,$au to order of magnitude \citep{Temmer2021}.

This model suggests that, by the time an exciter reaches a distance conducive to exciting plasma emission in the OVRO-LWA band, the material would still be in front of the star.
This conclusion may change if the flare that the material originated from occurred much closer to the limb.
There is indeed evidence of white-light flares from EK~Dra originating from the loop material rather than being dominated by footpoint emission, implying that the flare occurred on the limb or even partially behind it \citep{Namekata2022}.
However, such a flare produces a substantially different morphology than is observed here---the loop-dominant flare had a very low  amplitude ($0.3\%$) and persisted for hours whereas the flare evaluated here had an amplitude $10\times$ higher and only lasted $\sim10\,$min.
Differences in the instruments cannot adequately explain such a morphological difference.
Thus, it is likely that this flare did not occur very near the limb.

It is also worth noting that the size of the exciter responsible for the radio emission may be very large by the time it reaches the relevant distances for the OVRO-LWA---if we assume that emission originates from the entire exciter's surface, then the star may not substantially obscure the emission even if it originates behind the star.
Evaluating detectability in this case would require a more rigorous treatment of propagation effects on the emission, but it poses additional considerations for future flare-burst association endeavors.
These are: how much should we rely on flares for identifying radio emission if LOS effects alone may explain the presence of one and absence of the other? And how much will this inform future derivations of stellar flare-CME occurrence rates?
Observing burst signatures contemporaneous with flares remains important for understanding the coupling between thermal and non-thermal acceleration processes, but our understanding of the full flare-particle relationship may rely more on a statistical analysis of unrelated events.
\begin{figure}
    \centering
    \includegraphics[width=0.9\linewidth]{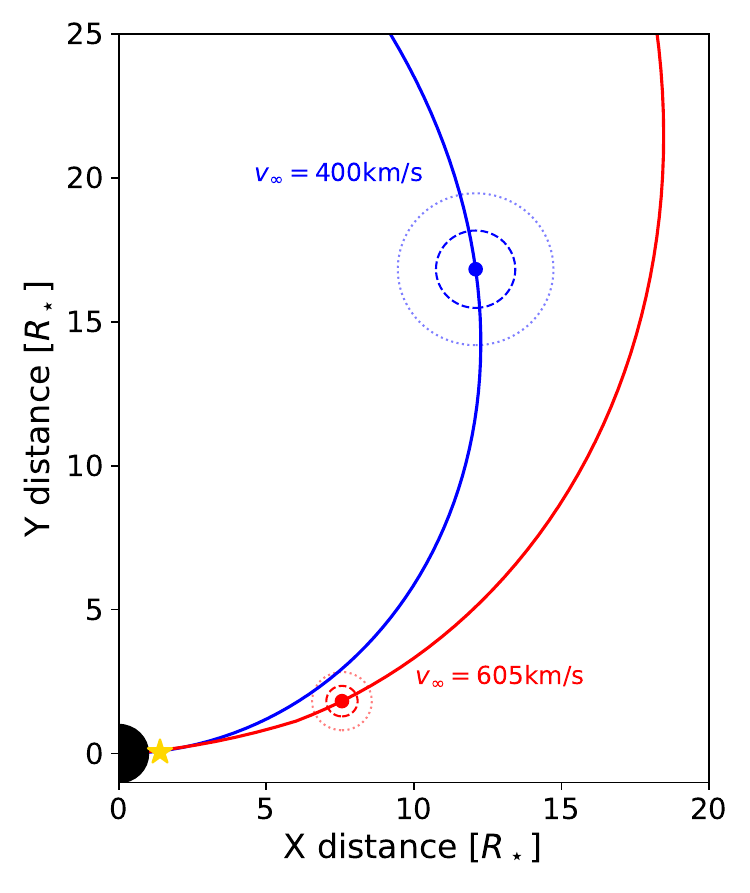}
    \caption{Spiral predictions for the CWS model in blue and the Parker model in red. The central locations of the exciter indicate the distance conducive to exciting 87\,MHz plasma emission in the respective models and are indicated by dots. Two different expansions rate for the exciter are demonstrated here: the dashed line indicates an expansion rate of 0.5 and the dotted line is an expansion rate of 1. The yellow star indicates the approximate, expected location of the reconnection event responsible for the flare and plasma acceleration.}
    \label{fig:spiral}
\end{figure}

}

\section{Summary}
We presented here the first season of of the SWAYS program which included 500\,hr of optical photometry; 66\% of these observations had complementary data at frequencies $<87\,$MHz, making this the most extensive coordinated observing program for solar-type stars in the radio and optical and the largest dedicated program to solar-type stars below 100\,MHz.
Our search for flares identified one unambiguous flare which originated from EK~Dra and had an energy consistent with superflares reported for the star in previous works.
Previous reports of \added{chromospheric} mass motion associated with flares of this energy had been considered possibly eruptive \added{on EK~Dra}, but we found no radio signatures that would indicate material escaping into EK~Dra's upper atmosphere.
This could imply that even these highly energetic flares are indeed not always accompanied by CMEs---this is in contrast to the solar flare-CME occurrence relation that shows solar CMEs almost always occur for flare energies $>10^{30}\,$erg. 
This in turn suggests that CMEs may not contribute as substantially to the bulk mass and angular momentum loss of stars as has been suggested.

Alternatively, there are a variety of conditions specific to EK~Dra's atmosphere that could \added{prevent} it from producing solar radio burst analogs\added{. It} is difficult to extrapolate solar expectations to stellar contexts without a better understanding of both the \added{specific stellar }environment and the conditions under which bursts occur \added{on} the Sun.
\added{Our analysis of the factors that may lead to a non-detection of radio emission benefited greatly from how extensively EK~Dra has been studied as a young solar analog.
X-ray, radio, and UV observations informed our treatment of its corona \citep{Gudel1995, Telleschi2005, Ayres2015}, magnetic field studies allowed us to estimate loop properties and the possibility of confinement \citep{Kochukhov2020, Namekata2024b}, and extensive monitoring of its flares and  CME signatures allowed us to contextualize this event relative to its chromospheric evidence of mass motion \citep{Namekata2022Nature, Namekata2024, Namekata2024b, Namekata2025TwoTemp}.
}

\added{Being able to use these properties revealed  previously unexplored limitations to burst detections at low frequencies. Perhaps the most salient are:
\begin{enumerate}
    \item {the very dense coronae of active stars require that CMEs and electron beams travel far from the stellar surface; evidence of bursts $\lesssim100\,$MHz would then essentially guarantee that mass has escaped into the interplanetary medium}
    \item {these large distances may also delay the occurrence of type~II bursts relative to the reconnection process by many hours}
    \item {the large distances and high coronal densities soften the super-Alfv\'enic requirement for shock development, even for $\approx1$kG surface fields}
    \item {the very high coronal temperatures may limit the development of type~III bursts.}
\end{enumerate}
Despite using properties specific to EK~Dra, the requirements for observing a burst change substantially depending on the wind model used. 
Furthermore, although the OVRO-LWA sensitivity may be sufficient to explain the non-detections---requiring the equivalent of a $\sim10^{10}\,$SFU type~III burst from the Sun for 10\,s integration time---solar scaling laws suggest that a type~II burst would have been detectable. 

}

\added{These results present a two-fold problem for studying space weather at low frequencies: more active stars flare more frequently and (might be) more likely to impulsively launch material, but the radio emission may be fully inhibited or else be delayed by hours. This makes it difficult to both coordinate observations across the spectrum as well as to associate low-frequency radio emission with the flare. 
Conversely, less active stars may facilitate the plasma instability more effectively  and allow for bursts that occur within minutes of the flare, but the highly-energetic flares that are assumed to be necessary for bright radio emission are much more rare than for active stars. 

Such a conclusion highlights the importance of monitoring and high-cadence surveys across the spectrum if we hope to identify both bursts and the flares presumed responsible for them.
Thus, a flare-agnostic radio-burst search with the OVRO-LWA may unveil bursts associated with magnetic activity below the energy threshold of optical instruments or which occur behind the limb of the star.
The all-sky field of the OVRO-LWA is complemented well by the large field of view of TESS in this respect. 
Such a coordination would also allow for investigating whether plasma emission from active stars does indeed occur long after their flare counterpart.
Although coordination of the OVRO-LWA and Flarescope may not reveal both the flare and resulting type~II burst in an extreme corona like EK~Dra's, it is still valuable for the less active stars in the SWAYS sample as well as for type~III burst searches.
Weather may limit the observing cadence of Flarescope, but it remains important as a dedicated monitoring tool. 
This is especially true for very nearby stars which saturate many other optical instruments.

Overall, the coordination of optical and radio wavelengths remains a potentially powerful tool to characterize space weather and stellar environments.
Using this tool effectively may require we rethink how we coordinate observations, what parameter space we should be exploring for de-dispersion searches, and how the extremity of stellar environments relative to the Sun's influences those decisions.
}

\section{Acknowledgments}
This material is based in part upon work supported by the National Science Foundation under grant number AST-1828784, the Simons Foundation (668346, JPG), the Wilf Family Foundation and Mt. Cuba Astronomical Foundation. A part of this work was carried out at the Jet Propulsion Laboratory and the California Institute of Technology under a contract with the National Aeronautics and Space Administration and funded through the JPL Researchers on Campus program.


\begin{appendix}

\section{Wind Profiles}\label{app:wind_profiles}
Constraining the properties of stellar winds is notoriously difficult. 
The goal here is not to accurately represent EK~Dra's environment, but to illustrate the ways in which coronal properties may impact \added{the search for} detect stellar bursts.
For this, we consider two, one-dimensional radial profiles for the wind of EK~Dra; one is representative of a Parker wind and the other assumes a constant wind speed (CWS) of 400\,km\,s$^{-1}$\added{. This latter speed} is typical for estimating stellar mass loss rates \citep{Wood2005, Fichtinger2017}. 
For both of these wind models, we assume that EK~Dra's magnetic field has a surface strength, $B_0$, of 1\,kG as informed by the 1.4\,kG field estimated by \citet{Kochukhov2020}.
We assume the field is dipolar such that $B = B_0 (r/R_{\star})^{-3}$. 
We also take its base coronal electron density, $n_0$, to be $4\times10^{10}\,\text{cm}^{-3}$, and assume its corona is isothermal with temperature $T_{\text{cor}} =10^7$; these values are derived from X-ray data in \citet{Gudel1995}.
We assume the corona is purely ionized Hydrogen such that the mass density $\rho=m_p\,n$ where $m_p$ is the proton mass and $n$ is the number density.

We consider the Parker model first, for which we follow a very similar prescription to \citet{Konijn2025}. 
We start by computationally solving for wind velocities, $v$, at distances $r$ that satisfy equation~17 from \citet{Parker1965}:
\begin{equation}
    \frac{v^2}{c_0^2} - \ln(\frac{v^2}{c_0^2}) = -3 + 4\,\ln(\frac{r}{r_c}) + 4\frac{r_c}{r}
\end{equation}
where $c_0 = {(k_B\,T_{\text{cor}}}/m_{\text{x}})^{1/2}$ is the isothermal sound speed of the corona and $r_c$ is the distance where the sound speed equals the orbital velocity.
By mass conservation, the mass density then scales like:
\begin{equation}
    \rho = \rho_0 \left(\frac{v}{v_0}\right)^{-1} \left(\frac{r}{R}\right)^{-2},
\end{equation}
where $\rho_0$ is the mass density at the base of the corona. 
Assuming that the ions and electrons move as a single fluid, then $\rho$ and $\rho_0$ can be replaced with $n$ and $n_0$, respectively.

At far distances, the ram pressure of the wind, $P_{\text{ram}} = 1/2 \rho\,v^2$, will substantially distort the magnetic field until it evolves effectively purely radially; the field strength then scales like $B\propto r^{-2}$.
In this highly simplified model where we ignore the effects of, e.g., rotation, the radial distance this occurs at is the same as the Alfv\'en radius, where the wind velocity exceeds the Alfv\'en velocity: $r_{\text{A}} = B\,(4\pi\rho_{\text{A}})^{1/2}$, where $\rho_{\text{A}}$ is the mass density at the distance $r_{\text{A}}$.
We assert that the wind reaches its terminal velocity, $v_\infty$, at this point so that the wind remains constant with distance beyond it and, consequently, the density scales purely $\propto r^{-2}$.

\begin{figure}
    \centering
    \includegraphics[width=0.6\linewidth]{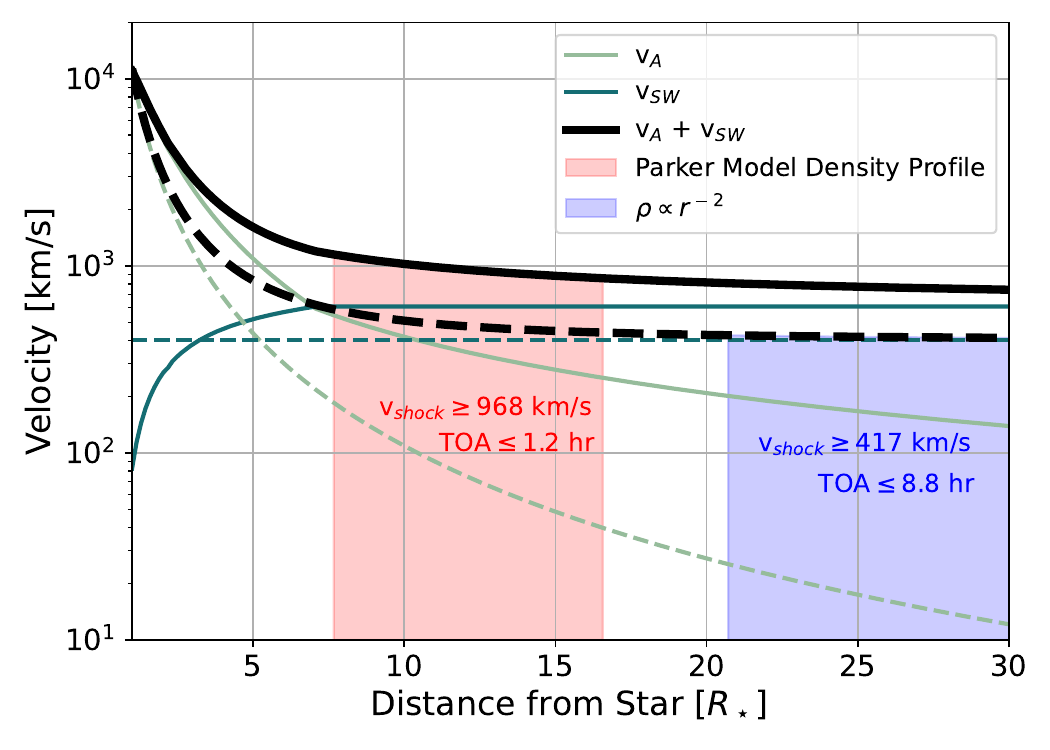}
    \caption{The wind velocity profiles for EK~Dra assuming a Parker model solution (solid lines) and a constant-velocity solution (dashed lines). The filled areas indicate the distances where the density is conducive to observing a type~II in the OVRO-LWA band; the red area indicates the range for the Parker wind solution and the blue area indicates the range for CWS model. $v_\text{shock}$ is the minimum speed required to drive a shock, reported in Table \ref{tab:wind_results}. }
    \label{fig:wind_profiles}
\end{figure}

In the simpler wind model, we take the wind velocity to be constant and thus $n \propto r^{-2}$ for all distances.
We also take the magnetic field to be closed everywhere so that $B\propto r^{-3}$ for all distances.
We assert that in order for a CME to produce a type~II burst, it must drive a super-Alfv\'enic shock in the frame of the wind, so that the CME speed must be $>v + v_{\text{A}}$.
The results of these models and shock requirements are shown in Figure~\ref{fig:wind_profiles}; their impacts on burst shape and arrival time to the band were illustrated in Figures~\ref{fig:type_ii_shape} and \ref{fig:burst_shape_prediction}. 
We summarize these models and the requirements they impose for an associated type~II burst to show up in the OVRO-LWA band in Table~\ref{tab:wind_results}.
Notably, the Parker wind model requires high CME speeds to develop a shock, but it reaches the OVRO-LWA band at distances twice as close to the stellar surface as the CWS model---timing of the burst would not have limited the OVRO-LWA if the CME were driving a shock.
Conversely, \added{the }CWS\added{ model} has a much easier shock condition, but it would take a substantial amount of time for a CME to reach the distances conducive to observing in the OVRO-LWA band; observing at higher frequencies would alleviate this problem.

\begin{deluxetable*}{ccc}
\tablecaption{Summary of wind models and type~II production requirements.}\label{tab:wind_results}
    \tablehead{
    \colhead{Property} & \colhead{Parker Model} & \colhead{CWS Model}
    }
    \startdata
    $v_\infty$ & 605\,km\,s$^{-1}$ & 400\,km\,s$^{-1}$\\
    Shock speed & \added{968}\,km\,s$^{-1}$ & \added{417}\,km\,s$^{-1}$\\
    \added{OVRO-LWA distance} &\added{7.7\,$R_\star$} & \added{20.7\,$R_\star$}\\
    OVRO-LWA TOA & \added{1.2}\,hr & \added{8.8}\,hr \\
    $B(r<r_\text{A})$ & $B_0\,(r/R_\star)^{-3}$ & $B_0\,(r/R_\star)^{-3}$ \\
    $B(r\geq r_\text{A})$ & $B_0\,(r/r_\text{A})^{-1}$ & $B_0\,(r/R_\star)^{-3}$ \\
    \enddata
    \tablecomments{The shock speed is the \added{average }speed a CME would need to travel to produce a super-Alfv\'enic shock \added{at the locations conducive to producing emission in the OVRO-LWA band}. The OVRO-LWA distance is the distance at which the density is low enough to excite frequencies $\leq87\,$MHz.}
\end{deluxetable*}

\section{Flare Loop Estimations}\label{app:flare_loop}
To approximate the volume of the flare loop, we take the heated area to be the area of the loop footpoints; this may be estimated from the signal measured by Flarescope.
For the third dimension, we use the loop scale height $l$ expression from \citet{Namekata2022}:
 \begin{equation}
     l \sim 1.64 \times 10 ^9 \left( \frac{\tau_\textrm{decay}}{100 \,\textrm{s}}\right)^{2/5} \left( \frac{E_\text{flare}}{10^{30}\,\text{erg}} \right)^{1/5} [\text{cm}].
 \end{equation}
 Using the flare values reported in Section~\ref{sec:flare_char} and Table~\ref{tab:flare_info}, we find that the length of the flare loop should be $\approx2.6\times10^{10}\,$cm (about 0.4\,$R_\star$, and similar to loop lengths derived for EK Dra previously in, e.g., \citet{Namekata2022, Namekata2024}).
 We take the maximum foot point size to be $L_\text{peak, bol}\,(\sigma_\text{SB}\,T_\text{flare}^4)^{-1}$ where $L_\text{peak, bol}$ is the peak bolometric luminosity assuming the flare temperature $T_\text{flare}$ of 10,000\,K.
 A peak flare luminosity of $7.8\times10^{31}$\,erg/s then implies an overall volume of $\approx3.6\times10^{30}\,\text{cm}^3$.
 These values can then be used to estimate the kinetic pressure of the CME and the relevant heights where confinement might occur.

It is worth noting that the energy available to a flare is limited by the magnetic energy contained within the flare loop, $E_B$.
Thus, we can use the energy of the flare to approximate the magnetic field strength using:
\begin{equation}\label{eq:field_strength}
    E_\text{flare}\sim E_B = \frac{(B/\text{G})^2}{8\pi}\frac{V}{\text{cm}^{3}}\,\text{erg},
\end{equation}
where $B$ is the field strength of the loop and $V$ is the volume that the loop fills.
If we use the volume we calculated here along with the overall flare energy in equation \ref{eq:field_strength}, this leads to an estimated field strength of $\approx500$\,G.
This is remarkably close to what the field strength would be at $0.4\,R_\star$ if the surface field strength is 1.4\,kG like is estimated by \citet{Kochukhov2020} and that it is a dipole---the values we calculate here may indeed be reasonable order-of-magnitude estimates.

\end{appendix}
\end{CJK*}
\end{document}